\documentclass[pra,onecolumn,floatfix,a4paper,superscriptaddress]{revtex4}
\usepackage{bm,color,graphicx,amsmath,txfonts}
\usepackage{here}
\usepackage{array}
\usepackage{graphicx}
\usepackage[colorlinks, citecolor=blue,linkcolor=blue]{hyperref}
\usepackage{braket}
\usepackage{dsfont}
\begin{document}

\makeatletter
\renewcommand{\@biblabel}[1]{\makebox[2em][l]{\textsuperscript{\textcolor{black}{\fontsize{10}{12}\selectfont[#1]}}}}
\makeatother

\let\oldbibliography\thebibliography
\renewcommand{\thebibliography}[1]{%
  \addcontentsline{toc}{section}{\refname}%
  \oldbibliography{#1}%
  \setlength\itemsep{0pt}%
}

\title{Mass-correction-induced enhancement of quantum correlations even beyond entanglement in the $e^{+}e^{-} \rightarrow J/\psi \rightarrow \Lambda(p\pi^{-}) \bar{\Lambda}(\bar{p}\pi^{+})$ process at the BESIII experiment under memory effects}

\author{Elhabib \surname{Jaloum} }
\address{LPTHE-Department of Physics, Faculty of Sciences, Ibnou Zohr University, Agadir 80000, Morocco}

\author{Omar Bachain}
\address{LPHE-Modeling and Simulation, Faculty of Sciences, Mohammed V University in Rabat, Rabat, Morocco}

\author{Mohamed \surname{Amazioug} }
\email{m.amazioug@uiz.ac.ma}
\address{LPTHE-Department of Physics, Faculty of Sciences, Ibnou Zohr University, Agadir 80000, Morocco}

\author{Nazek Alessa}
\address{Department of Mathematical Sciences, College of Science, Princess Nourah bint Abdulrahman University, P.O. Box 84428, Riyadh 11671, Saudi Arabia}

\author{Wedad R. Alharbi}
\address{Physics Department, College of Science, University of Jeddah, Jeddah, 23890, Saudi Arabia}

\author{Rachid Ahl Laamara}
\address{LPHE-Modeling and Simulation, Faculty of Sciences, Mohammed V University in Rabat, Rabat, Morocco}
\address{Centre of Physics and Mathematics, CPM, Faculty of Sciences, Mohammed V University in Rabat, Rabat, Morocco}

\author{Abdel-Haleem Abdel-Aty}
\affiliation{Department of Physics, College of Sciences, University of Bisha, Bisha 61922, Saudi Arabia}

\date{\today}

\begin{abstract}

In this work, we derive the bipartite density matrix for the $e^{+}e^{-} \rightarrow J/\psi \rightarrow \Lambda(p\pi^{-}) \bar{\Lambda}(\bar{p}\pi^{+})$ process at BESIII. We evaluate the impact of mass corrections and memory effects (within Markovian and non-Markovian regimes) on quantum correlations even beyond entanglement. The dependence of these quantum properties on the scattering angle $\varphi$ is analyzed, with a particular focus on the impact of mass corrections. By comparing massless and mass-corrected scenarios, we demonstrate that the inclusion of mass effects enhances the maximum violation of the Bell inequality. While the qualitative temporal behavior remains unchanged, mass corrections quantitatively modify the angular distribution and introduce additional extrema at $\varphi=0$ and $\varphi=\pi$, thereby strengthening non-local correlations without altering their fundamental dynamical origin. An examination of the hierarchy of quantum correlations in baryon-antibaryon systems yields partial confirmation: $\text{Bell Nonlocality} \subset \text{Steering} \subset \text{Entanglement} \subset \text{Discord}$. Additionally, our results show that classical correlations serve to mitigate the decoherence and the decay of quantum correlations. This interplay between classical and quantum correlations suggests practical applications in quantum information and provides a robust framework for investigating baryon-antibaryon interactions.

\end{abstract}
\maketitle
\section{Introduction}    \label{sec:1}

The conceptual foundations of quantum theory underwent a rigorous challenge in 1935 when Einstein, Podolsky, and Rosen (EPR) introduced their famous thought experiment aimed at exposing the theory's perceived incompleteness. The EPR paradox revealed the existence of non-local correlations, suggesting that the wave-function collapse induced by a measurement on one particle could instantaneously influence its distant counterpart \cite{ref1}. In response, Schr\"odinger introduced the pivotal concepts of entanglement and quantum steering to provide a theoretical framework for these intrinsic non-localities \cite{ref2}. Subsequently, Bell's theorem demonstrated that no theory based on local hidden variables could fully account for the non-local correlations observed between spatially separated systems during local quantum measurements \cite{ref3}. While entanglement is recognized as a uniquely quantum resource essential for various information-processing tasks \cite{ref4}, it is not the only relevant form of correlation. Both theoretical \cite{ref7} and experimental \cite{ref8} evidence suggest that other phenomena, such as quantum non-locality without entanglement, are equally significant for the development of quantum technologies \cite{ref5,ref6}, with certain separable states even outperforming their classical counterparts.

Historically, the manipulation of quantum information was primarily viewed through the lens of entanglement and separability \cite{ref8}. However, contemporary research has revealed that entanglement is not the sole requirement for implementing quantum protocols; indeed, some separable states can exhibit superior performance compared to classical systems \cite{ref9,ref10}. This realization has driven the creation of new metrics to identify non-classical correlations beyond entanglement, thereby expanding our control over quantum information \cite{ref11}. One such metric is quantum steering, which evaluates the asymmetric relationship between two observers, Alice and Bob. Within this framework, Alice can effectively "steer" or manipulate Bob’s state by utilizing their shared entanglement \cite{ref14}. Nevertheless, these non-classical correlations remain highly susceptible to environmental interference. Such interactions lead to decoherence, which destabilizes quantum states and facilitates the transition toward classical physical behavior \cite{ref15}.

It is now well established that quantum entanglement does not provide a complete characterization of all nonclassical correlations present in a composite quantum system \citep{refe1,refe2,Bachain26}. To address this limitation, Ollivier and Zurek introduced the concept of quantum discord (QD), which captures a broader class of quantum correlations extending beyond entanglement \citep{refe1}. In particular, QD can reveal nonclassical correlations even for separable mixed states, where entanglement is absent \citep{refe1,refe2}. Despite its conceptual significance, obtaining closed analytical expressions for this information–theoretic quantity is generally challenging. Consequently, several alternative correlation measures based on geometric or distance-type approaches have been developed, including geometric quantum discord \citep{refe3}, global geometric quantum discord \citep{refe4}, super quantum discord \citep{refe5}, trace-distance discord (TDD) \citep{refe6,refe7}, and quantum consonance \citep{refe8,refe9}.

The investigation of quantum entanglement has expanded into the realm of high-energy physics (HEP). Early studies examined these correlations in low-energy protons \citep{ref34}, while more recent work explores entanglement in hadronic final states at colliders \citep{ref35} and at increasingly smaller length scales \citep{ref36}. Experimental efforts to probe entanglement in HEP utilize a variety of systems, such as neutral kaons \citep{ref37, ref38, ref39, ref40}, positronium \citep{ref41}, neutral B-meson flavor oscillations \citep{ref42}, charmonium decays \citep{ref43}, and neutrino oscillations \citep{ref44}. A major milestone was recently reached with the observation of entanglement in top quark pair production at the LHC \citep{ref45}, alongside studies demonstrating the feasibility of testing Bell inequality violations in such systems \citep{ref46}. Current research continues to explore entanglement in top quark production \citep{ref47, ref48, ref49, ref50, ref51}, baryon production \citep{ref52}, and gauge boson sectors, specifically those arising from Higgs boson decays or direct production processes \citep{ref53, ref54, ref55, ref56}.

The discrete symmetries $P$ and $CP$ in the coupling of charmonium states to baryon–antibaryon pairs have been proposed as promising probes in high–statistics experiments \cite{He:1992ng,He:1993ar,He:2022jjc,Du:2024jfc}. This possibility is particularly relevant in view of the tens of billions of $J/\psi$ events already collected by the BESIII experiment, as well as the significantly larger data samples anticipated at the future Super Tau-Charm Facility (STCF). Such large statistics open the way to a determination of the weak mixing angle and to stringent constraints on the electric dipole moments of octet baryons at the charmonium resonance energies \cite{Du:2024jfc,Chen:2025rab}. The complete angular distributions describing the full decay chain can be expressed within the helicity formalism \cite{Zhang:2010zzo,Zhang:2009at,Fu:2023ose,Ovsiannikov:2025gcy}, which has been employed by the BESIII Collaboration to investigate baryon electric dipole moments induced by $CP$-violating effects \cite{BESIII:2025vxm}.

In this work, we derive the bipartite density matrix for the $e^{+}e^{-} \rightarrow J/\psi \rightarrow \Lambda(p\pi^{-}) \bar{\Lambda}(\bar{p}\pi^{+})$ process at BESIII. we explore several manifestations of quantum correlations, including Bell nonlocality, quantum steering, quantum concurrence, and quantum discord, in the reaction 
$e^{+}e^{-} \rightarrow J/\psi \rightarrow \Lambda(p\pi^{-})\,\bar{\Lambda}(\bar{p}\pi^{+})$ 
studied at the BESIII experiment. Here, $\Lambda$ and $\bar{\Lambda}$ denote a spin-$\frac{1}{2}$ baryon and its corresponding antibaryon. The analysis relies on the two-qubit density matrix describing the $\Lambda\bar{\Lambda}$ pair. We first examine the impact of baryon mass effects by comparing the massless approximation with a formulation that incorporates mass-dependent contributions. Our results show that taking these corrections into account enhances the maximal violation of the Bell inequality, leading to a larger extremal value of the Bell parameter than in the massless limit. Furthermore, the behavior of the considered quantum correlations exhibits a pronounced dependence on the scattering angle $\varphi$. We also study the decoherence dynamics of the $\Lambda\overline{\Lambda}$ system when it interacts with a classically correlated dephasing channel. In this context, classical correlations are assumed to exist between successive actions of the channel on the baryon and the antibaryon. We demonstrate that such correlations in the environmental noise can effectively mitigate the degradation of quantum correlations.

The paper is organized as follows. Sections~\ref{sec:2} and \ref{sec:3} introduce the physical model of the $\Lambda\bar{\Lambda}$ system and review the quantum correlation measures considered here, including Bell nonlocality, quantum steering, concurrence, and discord. Sections~\ref{sec:4} and \ref{sec:5} present the hierarchy of correlations for massless and mass-corrected scenarios and analyze the system dynamics under a correlated dephasing channel in both Markovian and non-Markovian regimes, respectively; the main conclusions are summarized in Sec.~\ref{sec:6}.

\section{Spin density matrix} \label{sec:2}

The state of a spin-$\frac{1}{2}$ particle pair (e.g., $\Lambda\bar{\Lambda}$) is generally given by the following density matrix
\begin{equation}
\rho_{{\Lambda\bar{\Lambda}}} = \frac{1}{4}\sum_{\mu,\nu}^{3}\hat{C}_{\mu\nu}\sigma^{\Lambda}_{\mu} \otimes \sigma^{\bar{\Lambda}}_{\nu},
\label{eq:1}
\end{equation}
The formalism employs a set of four Pauli matrices, $\sigma^{\Lambda}_{\mu}$ and $\sigma^{\bar{\Lambda}}_{\nu}$, defined in the rest frames of the baryon and antibaryon, with $\hat{C}_{\mu,\nu}$ representing a $4\times 4$ real matrix encoding spin correlations and polarizations. The spin operators are expressed in coordinate frames with axes $\mathbf{\hat{x}_1},\mathbf{\hat{y}_1},\mathbf{\hat{z}_1}$ and $\mathbf{\hat{x}_2},\mathbf{\hat{y}_2},\mathbf{\hat{z}_2}$ for the baryon and antibaryon, respectively.
\begin{figure}[!h]
\begin{center}
\includegraphics[width=8.5cm,height=6cm]{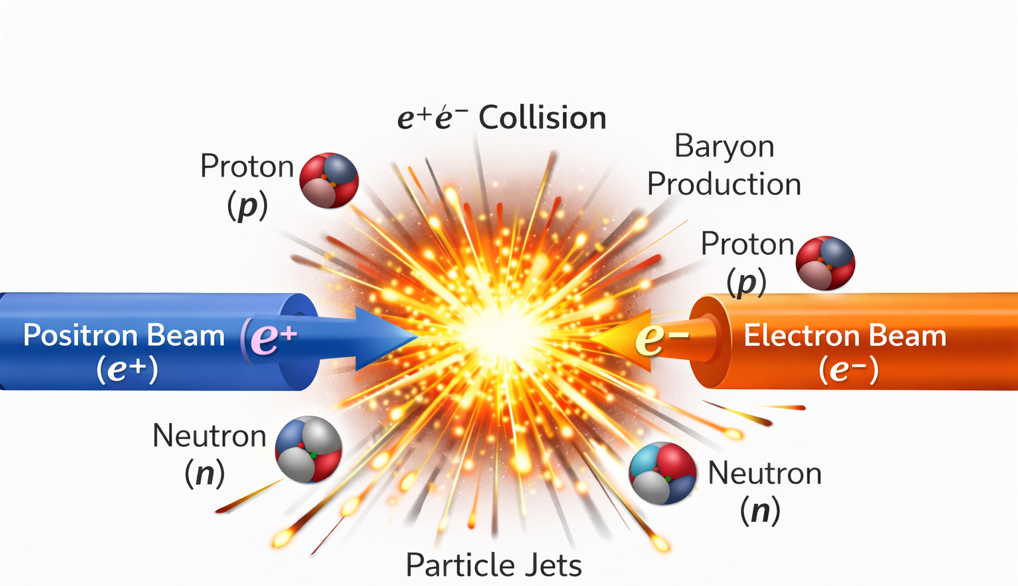}
\includegraphics[width=8.5cm,height=6cm]{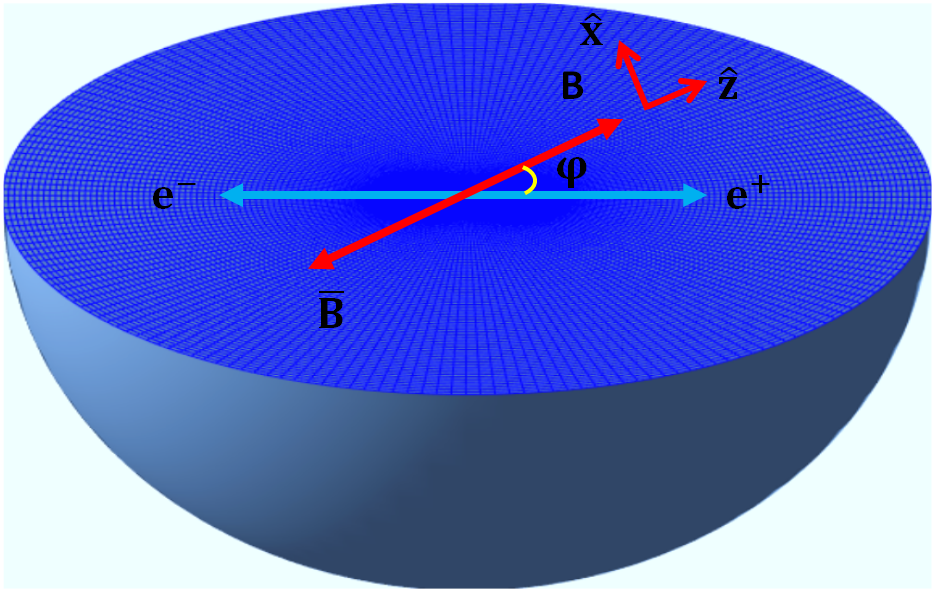}
\put(-485,160){{\bfseries (a)}}
\put(-240,160){{\bfseries (b)}}
\end{center}
\caption{(a) Schematic representation of the process $e^{+}e^{-} \to J/\psi \to \Lambda(p\pi^{-}),\bar{\Lambda}(\bar{p}\pi^{+})$. (b) The orientation of the coordinate axes $\{\mathbf{\hat{x}}, \mathbf{\hat{y}}, \mathbf{\hat{z}}\}$, established in the mutual rest frame of the $\Lambda$ and $\bar{\Lambda}$ pair.}
\label{fig:ee}
\end{figure}

In the $e^+ e^-$ center-of-momentum system, the $\hat{C}_{\mu\nu}$ matrix elements depend explicitly on the $\Lambda$ production angle. Only eight coefficients $\hat{C}_{\mu,\nu}$ are nonzero, and they are given by \citep{Mod1,Mod2,Mod3,Mod4,Mod5,Mod6}
\begin{equation}
\hat{C}_{\mu \nu}= \left(
\begin{array}{cccc}
\hat{C}_{0,0} & \hat{P}_x & \hat{P}_y    & \hat{P}_z \\
\hat{P}_{x} & \hat{C}_{x,x} & \hat{C}_{x,y} &  \hat{C}_{x,z} \\
\hat{P}_{y} & \hat{C}_{y,x} & \hat{C}_{y,y} & \hat{C}_{y,z} \\
\hat{P}_{z} & \hat{C}_{z,x} & \hat{C}_{z,y} & \hat{C}_{z,z}  \\
\end{array}
\right) \,,\label{eq:Omega}
\end{equation}

The parameters in Eq. (\ref{eq:Omega}) are given by $\hat{C}_{0,0} = 1$, $\hat{C}_{i,0} = \hat{P}_i^{+}$, $\hat{C}_{0,j} = \hat{P}_j^{-}$, and $\hat{C}_{i,j} = \hat{P}_{i,j}$. Here, $\hat{P}_{k}^{\pm}$ represents the polarization of the $\Lambda$ or $\bar{\Lambda}$ particle along the $\hat{\boldsymbol{\rm x}}, \hat{\boldsymbol{\rm y}}, \hat{\boldsymbol{\rm z}}$ directions, which are defined relative to the production plane. We decompose $\hat{C}_{\mu,\nu}$ into 12 matrices as follows \citep{Mod6}
\begin{equation}
\begin{aligned}
\hat{C}_{\mu,\nu} = \sum_{\varepsilon}\hat{C}_{\mu\nu}(\rm \varepsilon) + \sum_{\varepsilon}\hat{C}^{\text{(m)}}_{\mu\nu}(\varepsilon)
\label{eq:CmunvFinal}
\end{aligned}
\end{equation}
The spin density matrices $\hat{C}_{\mu,\nu}(\varepsilon)$, with $\varepsilon = {EM}, {AT}, {AE}, {TM}, {AM}, {TE}$, represent, respectively, the combinations of electric and magnetic form factors, $P$- and $CP$-violating strengths, $P$ with electric, $CP$ with magnetic, $P$ with magnetic, and $CP$ with electric form factors, while superscripts $(m)$ indicate mass corrections, and the matrices are systematically classified according to these distinct form factor combinations.

The quantum correlations of the spin density matrix $\hat{C}_{\mu,\nu}(EM)$ have already been investigated in \cite{JA1} and quantum teleportation was studied in \citep{JA2}.
In the present work, we restrict our study to the matrix $\hat{C}_{\mu,\nu}({AT})$ and $\hat{C}^{\rm (m)}_{\mu,\nu}({AT})$.

A set of four Pauli matrices, $\sigma^{\Lambda}_{\mu}$ ($\sigma^{\bar{\Lambda}}_{\nu}$), is defined in the rest frames of the baryon $\Lambda$ and antibaryon $\bar{\Lambda}$, with $\hat{C}_{\mu,\nu}$ representing a $4\times 4$ real matrix of their polarization and spin correlations, and the helicity rest frame of $\Lambda$ is defined as illustrated in Fig.~\ref{fig:ee}(b).
\begin{equation}
\hat{\boldsymbol{\rm y}} = \frac{\hat{\boldsymbol{\rm P}}_e \times \hat{\boldsymbol{\rm P}}_\Lambda}{|\hat{\boldsymbol{\rm P}}_e \times \hat{\boldsymbol{\rm P}}_\Lambda |}, \quad \hat{\boldsymbol{\rm z}} = \boldsymbol{\hat{\rm P}}_\Lambda, \quad \boldsymbol{\hat{\rm x}} = \boldsymbol{\hat{\rm y}} \times \boldsymbol{\hat{\rm z}}.
\label{eq:haty}
\end{equation}

The matrix elements $\hat{C}_{i,j}$ are functions of the production angle $\varphi$ between the incoming electron and outgoing baryon momenta, with $\cos(\varphi) = \boldsymbol{\hat{\rm P}}_e \cdot \boldsymbol{\hat{\rm P}}_\text{B}$. 

The spin density matrix $\hat{C}_{\mu,\nu}(AT)$, using the coordinate system defined in Eq. (\ref{eq:haty}), of baryon-antibaryon pair correspond to the $P$ and $CP$ violating strengths

\begin{equation}
\hat{C}_{\mu,\nu}(AT)=\frac{1}{1-\alpha_{\psi}\cos^2\varphi}
\begin{pmatrix}
1-\alpha_{\psi}\cos^2\varphi & 0 & \gamma_{\psi} \sin\varphi \cos\varphi & 0 \\
0 & \sin^2\varphi & 0 & \beta_{\psi} \sin\varphi \cos\varphi \\
\gamma_{\psi} \sin\varphi \cos\varphi & 0 & \alpha_{\psi} \sin^2\varphi & 0 \\
0 & \beta_{\psi} \sin\varphi \cos\varphi & 0 & -\alpha_{\psi}+\cos^{2}\varphi
\end{pmatrix},
\label{M:1}
\end{equation}
and a mass-correction term $\hat{C}^{(m)}_{\mu,\nu}$ arising from the finite electron mass
\begin{equation}
\hat{C}^{(m)}_{\mu,\nu}(AT)=\frac{1}{1+\alpha_{\psi}\cos2\varphi}
\begin{pmatrix}
1+\alpha_{\psi}\cos2\varphi & 0 & \gamma_{\psi} \sin2\varphi & 0 \\
0 & \alpha_{\psi}+\cos2\varphi & 0 & \beta_{\psi} \sin2\varphi \\
\gamma_{\psi} \sin2\varphi & 0 & 1+\alpha_{\psi}\cos2\varphi & 0 \\
0 & \beta_{\psi} \sin2\varphi & 0 & -\alpha_{\psi}-\cos2\varphi
\end{pmatrix},
\label{M:2}
\end{equation}
The vector charmonium decay parameter is $\alpha_\psi \in [-1,1]$, with $\beta_{\psi} = \sqrt{1-\alpha_{\psi}^{2}} \, \sin\Delta\Phi$ and $\gamma_{\psi} = \sqrt{1-\alpha_{\psi}^{2}} \, \cos\Delta\Phi$, where $\Delta\Phi \in [-\pi,\pi]$ denotes the relative form factor phase. Transforming the two-qubit state from Eqs.~(\ref{eq:1}), (\ref{M:1}), and (\ref{M:2}) into an X-state by swapping the $\hat{\boldsymbol{\rm y}}$ and $\hat{\boldsymbol{\rm z}}$ axes and diagonalizing $C_{i,j}$ greatly simplifies the analysis \cite{X1}, yielding the resulting spin density matrix

\begin{equation} \label{varrhoX}
\rho_{\Lambda\bar{\Lambda}}^{\rm X} = \frac{1}{4} \left(\hat{\rm I} \otimes \hat{\rm I} + \hat{C}_{0,z}\Big( \sigma_z \otimes \hat{\rm I} + \hat{\rm I} \otimes \sigma_z\Big) + \sum_{i=1}^{3} \hat{C}_{i,i} \sigma_i \otimes \sigma_i \right).
\end{equation}
The matrix $\hat{C}_{\mu,\nu}$ for this state becomes
\begin{equation}
\hat{C}_{\mu\nu} =
\begin{pmatrix}
1 & 0 & 0 & \mathcal{A} \\
0 & \mathcal{B}_{1} & 0 & 0 \\
0 & 0 & \mathcal{B}_{2} & 0 \\
\mathcal{A} & 0 & 0 & \mathcal{B}_{3}
\end{pmatrix},
\label{eq:sp} \quad \text{where} \quad
\begin{aligned}
\mathcal{A} &= \frac{\gamma_{\psi}\cos(\varphi)\sin(\varphi)}{1-\alpha_{\psi} \cos^2(\varphi)}, \quad \mathcal{B}_{3} = \frac{\alpha_{\psi}\sin^2(\varphi)}{1 - \alpha_{\psi} \cos^2(\varphi)},\\  
\mathcal{B}_{1} &= \frac{1 - \alpha_{\psi} + \sqrt{\bigg(\alpha_{\psi} -\cos^22\varphi\bigg)^2 +\beta^{2}_{\psi}\sin^22\varphi}}{2(1-\alpha_{\psi}\cos^2(\varphi))}, \\  
\mathcal{B}_{2} &= \frac{1 - \alpha_{\psi} - \sqrt{\bigg(\alpha_{\psi} -\cos^22\varphi\bigg)^2 +\beta^{2}_{\psi}\sin^22\varphi}}{2(1-\alpha_{\psi}\cos^2(\varphi))},
\end{aligned}
\end{equation}
and the matrix of a mass-correction term $\hat{C}^{(m)}_{\mu,\nu}$ of for this state becomes
\begin{equation}
\hat{C}^{(m)}_{\mu,\nu} =
\begin{pmatrix}
1 & 0 & 0 & \mathcal{A} \\
0 & \mathcal{B}_{1} & 0 & 0 \\
0 & 0 & \mathcal{B}_{2} & 0 \\
\mathcal{A} & 0 & 0 & 1
\end{pmatrix},
\label{eq:sp} \quad \text{where} \quad
\begin{aligned}
\mathcal{A} &= \frac{\gamma_{\psi}\sin(2\varphi)}{1+\alpha_{\psi} \cos(2\varphi)}, \\  
\mathcal{B}_{1} &= \frac{\sqrt{\big(\alpha_{\psi}+\cos2\varphi\big)^2 +\beta^{2}_{\psi}\sin^22\varphi}}{1+\alpha_{\psi}\cos2\varphi}, \\  
\mathcal{B}_{2} &=-\frac{\sqrt{\big(\alpha_{\psi}+\cos2\varphi\big)^2 +\beta^{2}_{\psi}\sin^22\varphi}}{1+\alpha_{\psi}\cos2\varphi}.
\end{aligned}
\end{equation}

By applying Eq. (\ref{varrhoX}), the bipartite spin density operator can be explicitly reformulated in terms of the $\sigma_z$ eigenbasis. We rewrite the spin density operator for the baryon-antibaryon system in the $\sigma_{z}$ basis 
\begin{equation}
\rho_{\Lambda\bar{\Lambda}}^{\text{X}} =
\begin{pmatrix}
\rho_{1,1}& 0 & 0 & \rho_{1,4} \\
0 & \rho_{2,2} & \rho_{2,2} & 0 \\
0 & \rho_{2,2} & \rho_{2,2}& 0 \\
\rho_{1,4} & 0 & 0 & \rho_{4,4}
\end{pmatrix}, \quad \text{where} \quad
\begin{aligned}
\rho_{1,1}&= \frac{1}{4}\bigg(1+2\mathcal{A}+\mathcal{B}_{3}\bigg), \quad
\rho_{1,4} &= \rho_{4,1} = \frac{1}{4}\bigg(\mathcal{B}_{1}-\mathcal{B}_{2}\bigg), \\
\rho_{2,2} &= \rho_{3,3} = \frac{1}{4}\bigg(1-\mathcal{B}_{3}\bigg), \quad
\rho_{2,3} &= \rho_{3,2} = \frac{1}{4}\bigg(\mathcal{B}_{1}+\mathcal{B}_{2}\bigg), \\
\rho_{4,4} &= \frac{1}{4}\bigg(1-2\mathcal{A}+\mathcal{B}_{3}\bigg).
\end{aligned}
\label{eq:varrho}
\end{equation}
The spin density matrix $\hat{C}_{\mu,\nu}$, using the coordinate system defined in Eq. (\ref{eq:haty}), of baryon-antibaryon pair correspond to the P violating strength and electric form factor.

\section{Quantum resources} \label{sec:3}

This section is devoted to the study of the baryon-antibaryon spin density operator, an essential framework for characterizing the quantum structure of the $\Lambda\bar{\Lambda}$ system. From this density matrix, we quantify various quantum resources, including Bell non-locality, concurrence, geometric quantum discord, and quantum coherence.
\subsection{Bell non-locality}

Bell nonlocality stands as one of the most striking manifestations of quantum mechanics, capturing the fact that correlations between measurements performed on entangled particles elude any classical description \cite{b1,b2}. This phenomenon directly challenges local hidden variable theories, which attempt to account for such correlations through classical mechanisms \cite{b3}.

For bipartite systems of dimension $2 \times 2$, detecting nonlocality amounts to testing whether a given state $\rho$ violates the Bell-CHSH inequality, formulated by Clauser, Horne, Shimony, and Holt \cite{b3}. This inequality reads
\begin{equation}
    \vert\langle B_{\text{CHSH}}\rangle_{\rho}\vert \leq 2,
    \label{eq:bell_inequality}
\end{equation}
where $\langle B_{\text{CHSH}} \rangle_{\rho} = \text{tr}[\rho B_{\text{CHSH}}]$ denotes the expectation value of the Bell operator $B_{\text{CHSH}}$.

For any two-qubit state, the maximal achievable value of $\langle B_{\text{CHSH}} \rangle_{\rho}$, denoted $\mathtt{B}_{\max}(\rho_{\Lambda\bar{\Lambda}}) = \max |\langle B_{\text{CHSH}} \rangle_{\rho}|$, can be computed from the correlation matrix $X$ whose elements are given by $X_{\mu\nu} = {\rm Tr}(\rho\,\sigma_\mu \otimes \sigma_\nu)$. Letting $\omega_i$ ($i = 1,2,3$) be the eigenvalues of $X^\dagger X$, and defining $\mathcal{M}(\rho_{\Lambda\bar{\Lambda}}) = \max_{i<j}(\omega_i + \omega_j)$, one obtains the relation $\mathtt{B}_{\max}(\rho_{\Lambda\bar{\Lambda}}) = 2\sqrt{\mathcal{M}(\rho_{\Lambda\bar{\Lambda}})}$. A violation of inequality \eqref{eq:bell_inequality} occurs if and only if $\mathcal{M}(\rho_{\Lambda\bar{\Lambda}}) > 1$. Consequently, $\mathcal{M}(\rho_{\Lambda\bar{\Lambda}})$ provides a convenient quantifier of the degree of Bell nonlocality for a bipartite quantum state.

For the output state $\rho_{\Lambda\bar{\bar{\Lambda}}}$ introduced in Eq.~\eqref{eq:varrho}, one finds
\begin{equation}
    B_{\text{CHSH}}[\rho_{\Lambda\bar{\Lambda}}] = 2\sqrt{\max\{\mathcal{M}_1, \mathcal{M}_2\}},
    \label{eq:bell_output}
\end{equation}
where
\begin{align*}
    \mathcal{M}_1 &= 8\left(|\rho_{1,4}|^2 + |\rho_{2,3}|^2\right), \text{and} \quad
    \mathcal{M}_2 = 4\left(|\rho_{1,4}| + |\rho_{2,3}|\right)^2 + (\rho_{1,1} + \rho_{4,4}-2\rho_{2,2})^2.
\end{align*}

From these expressions, a normalized measure of Bell nonlocality can be constructed as
\begin{equation}
\mathtt{B}\big(\varrho_{\Lambda\bar{\bar{\Lambda}}}\big)=\max\Bigg[0,\frac{B_{\text{CHSH}}-2}{2\sqrt{2}-2}\Bigg].
\end{equation}

This quantity vanishes when the state satisfies the Bell-CHSH inequality, and increases monotonically with the degree of violation, reaching its maximum at the Cirel'son bound $2\sqrt{2}$.
\subsection{Quantum steering}
Quantum steering, a phenomenon that captures non-local correlations intermediate between entanglement and Bell nonlocality, can be characterized through the violation of specific steering inequalities. In this study, we employ the CJWR steering inequality, introduced by Cavalcanti, Jones, Wiseman, and Reid \citep{ST1, ST2}, to diagnose the steerability of the $\Lambda\bar{\Lambda}$ system. This criterion is particularly effective for two-qubit states subjected to three orthogonal measurement settings on each side ($N=3$). The steering function is defined as:

\begin{equation}
F_{3}^{\rm CJWR}(\rho)=\frac{1}{\sqrt{3}}\Bigg|\sum_{i=1}^{3}{\rm Tr}\left[ \rho\left( A_{i}\otimes B_{i}\right) \right] \Bigg|\leq 1,
\label{eq:CJWR}
\end{equation}

where $A_{i}=\mathbf{\hat{u}}_{i}\cdot\boldsymbol{\sigma}$ and $B_{i}=\mathbf{\hat{v}}_{i}\cdot\boldsymbol{\sigma}$ represent the projection operators for Alice and Bob, respectively, with $\boldsymbol{\sigma}$ being the vector of Pauli matrices. The unit vectors $\mathbf{\hat{u}}_{i}$ and orthonormal vectors $\mathbf{\hat{v}}_{i}$ define the local measurement directions.

The steerability of the state $\rho$ is fundamentally governed by its $3\times 3$ correlation matrix $C$, with elements $C_{ij} = {\rm Tr}[\rho(\sigma_i \otimes \sigma_j)]$. To extract the maximum steering violation, the measurement axes must be optimized relative to the principal axes of $C$. By setting $\parallel C\mathbf{\hat{v}}_{i}\parallel = \sqrt{{\rm Tr}(CC^{\rm T})/3}$ and choosing $\mathbf{\hat{u}}_{i}$ to align with the steered states \citep{ST3}, the function $F_{3}^{\rm CJWR}$ reaches its maximum:

\begin{equation}
\mathcal{F}_{3}(\rho)=\max_{\mathbf{\hat{u}}_{i},\mathbf{\hat{v}}_{i}}\left[ F_{3}^{\rm CJWR}(\rho)\right]  =\sqrt{{\rm Tr}\left( C C^{\rm T}\right)}
\end{equation}

For the specific case of $\Lambda\bar{\Lambda}$ pairs, which are described by an X-shaped density operator due to parity and $CP$ conservation in the production process, the maximal violation simplifies to $\mathcal{F}_{3}(\rho_{\Lambda\bar{\Lambda}})=\sqrt{\mathcal{B}^{2}_{1}+\mathcal{B}^{2}_{2}+\mathcal{B}^{2}_{3}}$. Here, $\mathcal{B}_i$ correspond to the spin-correlation parameters along the principal axes.

Since a state is considered steerable only if $\mathcal{F}_{3} > 1$, and given that the theoretical maximum for a singlet state is $\mathcal{F}_{3}^{\max}=\sqrt{3}$, we introduce a modified measure $\mathtt{S}(\rho_{\Lambda\bar{\Lambda}})$ to quantify the degree of steering:

\begin{equation}
\mathtt{S}(\rho_{\Lambda\bar{\Lambda}})=\max\left\lbrace 0,\frac{\mathcal{F}_{3}(\rho_{\Lambda\bar{\Lambda}})-1}{\sqrt{3}-1}\right\rbrace 
\end{equation}

This normalized measure ensures that $\mathtt{S} = 0$ at the steering boundary and $\mathtt{S} = 1$ for maximally steerable states, providing a robust metric for comparing quantum correlations across different kinematic regions in baryon decay experiments.
\subsection{Concurrence}

Several entanglement measures exist to quantify the quantum correlations present in bipartite states, among which concurrence is one of the most widely used \citep{eg1,eg2}. For an arbitrary two-qubit state $\rho_{\Lambda\bar{\Lambda}}^{\rm X}$, concurrence is defined as

\begin{equation}
\mathtt{C}(\rho_{\Lambda\bar{\Lambda}}^{\rm X})=\max\left\lbrace  \sqrt{\mu_{1}}-\sqrt{\mu_{2}}-\sqrt{\mu_{3}}-\sqrt{\mu_{4}},0\right\rbrace,
\label{eq:C}
\end{equation}

where $\mu_i$ (with $i=1,2,3,4$) are the eigenvalues, arranged in decreasing order, of the auxiliary matrix

\begin{equation}
\mathcal{R}(\rho_{\Lambda\bar{\Lambda}})=(\sigma_{y}\otimes\sigma_{y})\,\rho_{\Lambda\bar{\Lambda}}^{\ast}\,(\sigma_{y}\otimes\sigma_{y}).
\end{equation}

Here, $\rho_{\Lambda\bar{\Lambda}}^{\ast}$ denotes the complex conjugate of $\rho_{\Lambda\bar{\Lambda}}$ in the computational basis, and $\sigma_y$ represents the Pauli matrix along the $y$-direction.

For the specific class of X-shaped density matrices considered in this work, the general expression simplifies considerably. In this case, concurrence takes the closed form

\begin{equation}\label{crx}
\mathtt{C}(\rho_{\Lambda\bar{\Lambda}}^{\rm X})=2\max \left\lbrace |\rho_{2,3}|-\sqrt{\rho_{1,1}\rho_{4,4}},\; |\rho_{1,4}|-\sqrt{\rho_{2,2}\rho_{3,3}},\;0\right\rbrace .
\end{equation}

The concurrence ranges from zero to one by construction. A vanishing concurrence indicates that the state is separable, meaning it exhibits no quantum entanglement. Conversely, a strictly positive concurrence signals the presence of entanglement, with higher values reflecting stronger quantum correlations between the two subsystems.
\subsection{Quantum discord}

Beyond the standard paradigm of entanglement, QD provides a more comprehensive measure of non-classical correlations in bipartite systems. Its primary advantage lies in its ability to capture quantum coherence even in separable states where entanglement—defined by the non-separability of the density matrix—is entirely absent \cite{Ollivier2001, Henderson2001}.

For an arbitrary bipartite state $\rho_{\Lambda\bar{\Lambda}}$, the total correlations are quantified by the quantum mutual information: \begin{equation} 
\mathcal{I}(\rho_{\Lambda\bar{\Lambda}}) = S(\rho_{\Lambda}) + S(\rho_{\bar{\Lambda}}) - S(\rho_{\Lambda\bar{\Lambda}}), 
\end{equation}
The von Neumann entropy is defined as $S(\rho) = -\mathrm{Tr}(\rho \log \rho)$. Total correlations in a bipartite system can be separated into classical and quantum contributions. The classical correlations, denoted $\mathcal{J}_{\Lambda}(\rho_{\Lambda\bar{\Lambda}})$, correspond to the maximal information about subsystem $\bar{\Lambda}$ that can be obtained through a projective measurement on subsystem $\Lambda$
\begin{equation} 
\mathcal{J}_{\Lambda}(\rho_{\Lambda\bar{\Lambda}})=\sup_{\left\lbrace B_{k}\right\rbrace }\left[ \mathcal{I}(\rho_{\Lambda\bar{\Lambda}}|{B_{k}})\right] . 
\label{eq:J} 
\end{equation} 
The quantum discord is then defined as the discrepancy between these two quantities: $\mathtt{QD}_{\Lambda}(\rho_{\Lambda\bar{\Lambda}}) = \mathcal{I}(\rho_{\Lambda\bar{\Lambda}}) - \mathcal{J}_{\Lambda}(\rho_{\Lambda\bar{\Lambda}})$. A non-zero discord signals the presence of quantum correlations that are intrinsically sensitive to, and perturbed by, the act of measurement.

In this work, we focus on a specific class of X-shaped states characterized by $\rho_{2,2} = \rho_{3,3}$ and real coherences. This structure leads to significant analytical simplifications: the reduced density matrices for subsystems $\Lambda$ and $\bar{\Lambda}$ become identical, $S(\rho_{\Lambda}) = S(\rho_{\bar{\Lambda}})$, rendering the classical correlations symmetric and independent of which subsystem is measured. 

Due to $CP$ conservation (as defined in Eq. \ref{eq:varrho}), the resulting baryon–antibaryon states possess an inherent symmetry. This property ensures that the quantum discord remains consistent regardless of whether the measurement is attributed to the baryon ($\Lambda$) or the antibaryon ($\bar{\Lambda}$).

An exact analytical framework exists for determining the QD of $X$-states \citep{QD1,QD2,QD3,QD4}, formulated as follows \begin{equation} 
\mathtt{QD}(\rho_{\Lambda\bar{\Lambda}}) = 1 - h\left(\frac{1+\mathcal{A}}{2}\right) + \sum_{i=1}^{4} \gamma_i \log_2 \gamma_i - \max_{\varepsilon \in [0,1]} F(\varepsilon) \label{eq:QD_general} 
\end{equation}
where $\lambda_i$ are the eigenvalues of the X-state density matrix $\rho_{\Lambda\bar{\Lambda}}^{\rm X}$ from Eq.~(\ref{eq:varrho}). For the class of states under study, these eigenvalues simplify to
\begin{equation}
\begin{aligned}
\gamma_{1,2} &= \frac{1}{2}\left[(\rho_{11}+\rho_{44}) \pm \sqrt{(\rho_{11}-\rho_{44})^2 + 4|\rho_{14}|^2}\right], \\
\gamma_{3,4} &= \rho_{22} \pm |\rho_{23}|,
\end{aligned}
\end{equation}

and $F(\varepsilon)$ is defined as
\begin{equation}
\begin{aligned}
F(\varepsilon)=&\frac{1+\mathcal{A}\varepsilon +\Delta_{+}}{4}\log_{2}\frac{1+\mathcal{A}\varepsilon +\Delta_{+}}{4}+\frac{1+\mathcal{A}\varepsilon -\Delta_{+}}{4}\log_{2}\frac{1+\mathcal{A}\varepsilon -\Delta_{+}}{4}\\
&+\frac{1-\mathcal{A}\varepsilon +\Delta_{-}}{4}\log_{2}\frac{1-\mathcal{A}\varepsilon +\Delta_{-}}{4}+\frac{1-\mathcal{A}\varepsilon -\Delta_{-}}{4}\log_{2}\frac{1-\mathcal{A}\varepsilon -\Delta_{-}}{4}
\end{aligned}
\end{equation}

where the optimization function $F(\varepsilon)$ is governed by the parameters
\[
\Delta_{\pm} = \sqrt{\mathcal{B}(1-\varepsilon^2) + (\mathcal{A} \pm \mathcal{B}_3 \varepsilon)^2},
\]
with $\mathcal{B} = \max\left\{ |\mathcal{B}_1|, |\mathcal{B}_2|\right\}$.

To determine the discord analytically, one must optimize a function $F(\varepsilon)$ over the interval $\varepsilon \in [0, 1]$. While this is generally a complex task, it has been established that for rank-2 mixed X-type states, the maximum is restricted to the boundaries $\varepsilon=0$ or $\varepsilon=1$ \cite{QD5}. Our numerical analysis confirms that for the $\Lambda\bar{\Lambda}$ pairs under investigation, the optimum is consistently achieved at $\varepsilon=0$. Consequently, the quantum discord reduces to a streamlined explicit relation \cite{QD6} 
\begin{equation} \mathtt{QD}(\rho_{\Lambda\bar{\Lambda}}) = h\left(\frac{1+\mathcal{A}}{2}\right) - h\left(\frac{1+\mathcal{B}_{3}}{2}\right) + h\left(\frac{1+\sqrt{\mathcal{B}_{1}^2 + \mathcal{B}_{3}^2 - \mathcal{B}_{1}\mathcal{B}_{3}}}{2}\right), 
\end{equation} 
where $h(x)$ represents the Shannon binary entropy. The coefficients $a$ and $\mathcal{B}_{1,2,3}$ are functions of the scattering angle $\varphi$, governed by the physical parameters $\alpha_{\psi}$ and $\Delta\Phi$. This result directly links fundamental quantum correlations to experimentally observable scattering parameters.
\section{Results and discussions}   
A comparative analysis is performed between the massless approximation and the theoretical framework incorporating mass-dependent corrections. Our results demonstrate that, while the fundamental dynamics of the process remain invariant, the inclusion of mass effects quantitatively modifies the correlation amplitudes. Specifically, mass corrections enhance the violation of Bell inequalities and induce distinct modulations in the angular distribution of observables. These findings underscore the necessity of integrating kinematic effects for a rigorous assessment of entanglement and quantum resources in high-energy baryon-antibaryon pairs.
\begin{table}[H]
\begin{center}
\caption{We use the numerical parameters with uncertainties for $e^{+}e^{-} \rightarrow J/\psi \rightarrow \Lambda(p\pi^{-}) \bar{\Lambda}(\bar{p}\pi^{+})$ as reported by the BESIII collaboration \cite{Tab}.}
\begin{tabular}{c c c c c c} 
\hline
\hline
& $\alpha_{\psi}$ & $\beta_{\psi}$ & $\gamma_{\psi}$ & $\Delta\Phi/rad$  \\
\hline
 $J/\psi \rightarrow \Lambda(p\pi^{-}) \overline{\Lambda}(\bar{p}\pi^{+})$  & $-0.32\pm 1.23$ & $0.85\pm 0.51$ & $-0.41\pm 0.72$ & $-4.26\pm 0.82$ \\
\hline
\label{tab1}
\end{tabular}
\end{center}
\end{table}
\subsection{Steady state} \label{sec:4}
\begin{figure}[!h]
\includegraphics[scale=0.4]{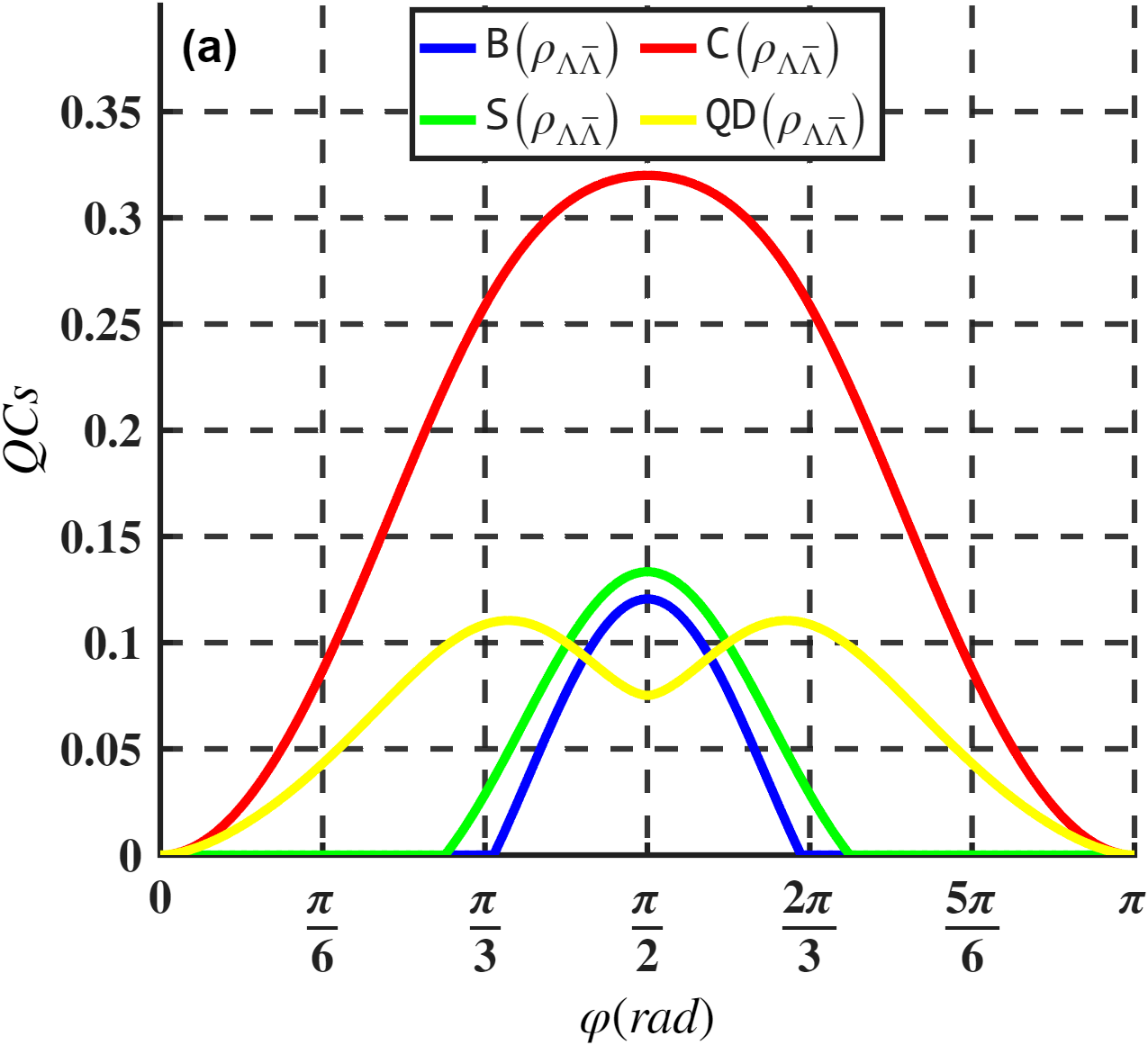}
\includegraphics[scale=0.4]{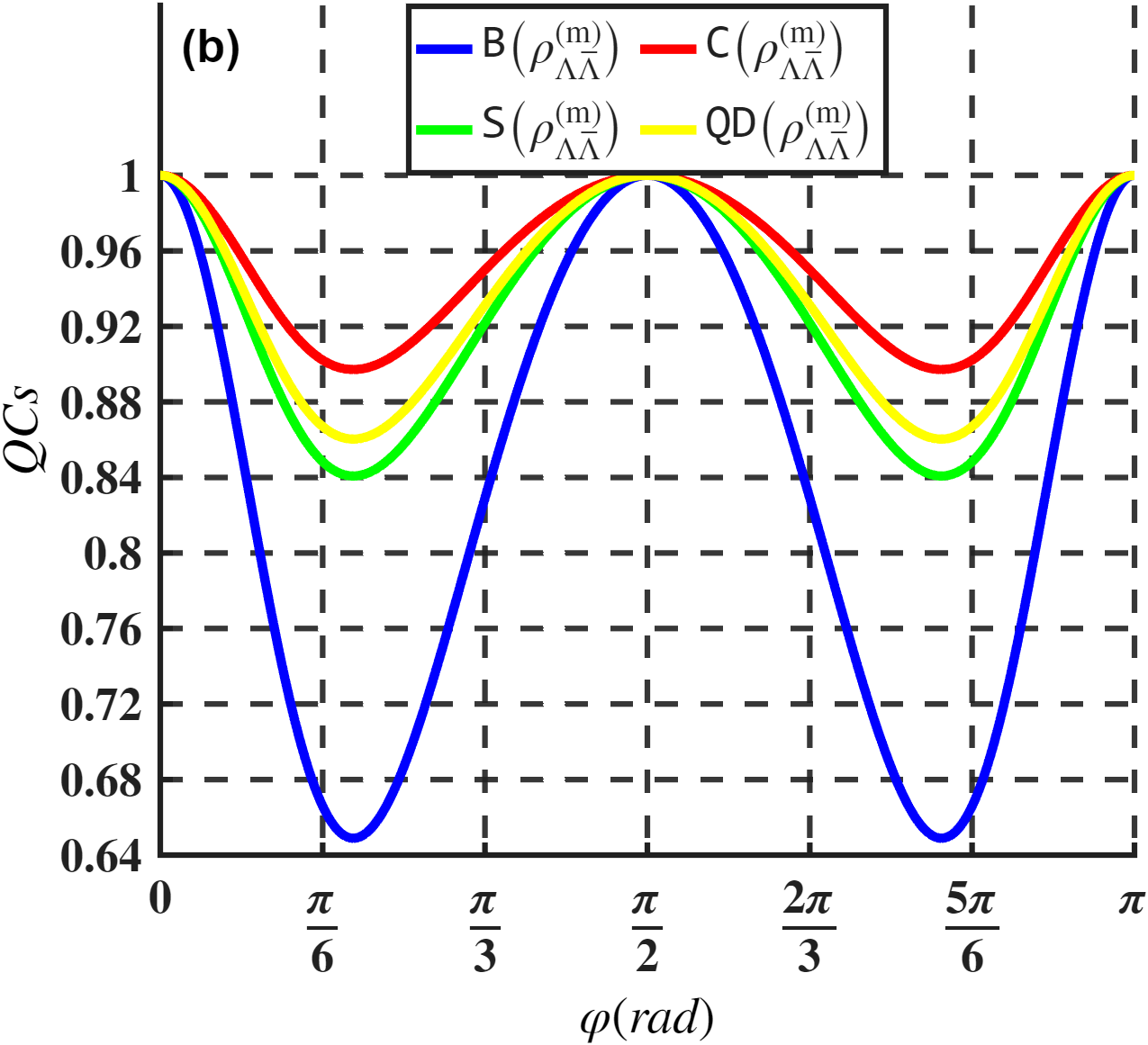}
\caption{We investigate four distinct types of quantum correlations—Bell nonlocality, steering, concurrence, and quantum discord—as functions of the scattering angle $\varphi$ in the process $e^{+}e^{-} \rightarrow J/\psi \rightarrow \Lambda(p\pi^{-}) \overline{\Lambda}(\bar{p}\pi^{+})$. The results are presented for two cases: (a) without mass corrections and (b) in the presence of mass corrections.}
\label{fig:QCs}
\end{figure}

In Fig.\ref{fig:QCs}, we compare five quantum correlation quantifiers-Bell non-locality ($\mathtt{B}$), quantum steerability ($\mathtt{S}$), entanglement ($\mathtt{C}$) and quantum discord ($\mathtt{QD}$)-as a function of the scattering angle $\varphi$ in the process $e^{+}e^{-} \rightarrow J/\psi \rightarrow \Lambda(p\pi^{-}) \overline{\Lambda}(\bar{p}\pi^{+})$. Figures 2(a) and 2(b) correspond to the $\Lambda\bar{\Lambda}$ baryon-antibaryon system without and with mass corrections, respectively.

Using the experimental parameters from Table~\ref{tab1}, the main objective of this study, as illustrated in Fig.~\ref{fig:QCs}(a–b), is to analyze Bell nonlocality, quantum steering, concurrence, and quantum discord for the $\Lambda\bar{\Lambda}$ decay channel.

This figure shows that these four types of correlations exhibit symmetry with respect to $\varphi = \pi/2$ within the interval $\varphi \in [0, \pi]$. As shown in Fig.~\ref{fig:QCs}(a), when the scattering angle $\varphi$ increases from $0$ to $\pi/2$, Bell nonlocality, quantum steering, and concurrence increase monotonically from zero. After reaching their maximum values at $\varphi = \pi/2$, these quantities then decrease monotonically and vanish at $\varphi = \pi$, as also depicted in Fig.~\ref{fig:QCs}(a).

It is also worth noting that the baryon-antibaryon system exhibits stronger quantum correlations as the scattering angle approaches $\varphi = \pi/2$, highlighting the crucial role of this angle in high-energy processes. This behavior emphasizes the significant impact of the scattering angle on the degree of entanglement in the $\Lambda\bar{\Lambda}$ system.

In contrast, although quantum discord also displays symmetry around $\varphi = \pi/2$, it remains nonzero over almost the entire range $\varphi \in [0, \pi]$, except for the collinear configurations at $\varphi = 0$ and $\varphi = \pi$. Moreover, its maximum value is not necessarily achieved at $\varphi = \pi/2$, unlike Bell nonlocality, quantum steering, and concurrence. This result highlights that even in the absence of maximal entanglement or nonlocality, discord-type quantum correlations persist, revealing the significant influence of quantum fluctuations in $\Lambda\bar{\Lambda}$ pairs.

In the presence of mass, we note that the correlations become stronger and can reach the value 1 for $\varphi = 0$, $\pi/2$, and $\pi$, as shown in Fig.~\ref{fig:QCs}(b). This amplification can be explained by the influence of mass corrections on the structure of the spin states, which can enhance the measurable quantum correlations, particularly at angles where the symmetry of the system is maximal.
In the study of baryon-antibaryon systems, quantum steering and quantum discord can be analyzed alongside entanglement and Bell non-locality. While these latter two forms of correlations have already been the subject of extensive investigation \cite{Hr1,Hr2}, it is essential to note that they all belong to a hierarchical structure inherent to quantum information theory. For any bipartite system, these correlations indeed satisfy the following inclusion order \cite{Hr3,Hr4}

\begin{equation}
\text{Bell non-locality}\subseteq\text{Steering}\subseteq\text{Entanglement}\subseteq\text{Discord}
\label{eq:Hr}
\end{equation}

In Fig.\ref{fig:QCs}(a), the amplitudes of Bell non-locality, quantum steerability, and entanglement exhibit a clear hierarchy: \[\mathtt{B}(\rho_{\Lambda\bar{\Lambda}})\subseteq \mathtt{S}(\rho_{\Lambda\bar{\Lambda}})\subseteq\mathtt{C}(\rho_{\Lambda\bar{\Lambda}})\] This indicates that the degree of entanglement is higher than that of steerability, which in turn is higher than that of Bell non-locality. In contrast, the amplitudes of quantum discord and purity do not show a clear hierarchical relationship with these three correlations in this specific context.

As shown in Fig.\ref{fig:QCs}(a), it is evident that quantum discord, entanglement, and purity remain non-zero across the entire range of the scattering angle (excluding the endpoints $0$ and $\pi$). Conversely, steerability is limited to a narrower range centered around $\varphi = \pi/2$, while Bell non-locality is even more constrained, situated entirely within the steerability region.

Relevant parameters associated with these correlations are listed in Table \ref{tab1}. By combining the results from Fig.\ref{fig:QCs} and Table \ref{tab1}, we observe that these correlations follow the global hierarchical relationship: \[\mathtt{B}(\rho_{\Lambda\bar{\Lambda}})\subset \mathtt{S}(\rho_{\Lambda\bar{\Lambda}})\subset\mathtt{C}(\rho_{\Lambda\bar{\Lambda}})\subset\mathtt{QD}(\rho_{\Lambda\bar{\Lambda}})\]. This implies that steerability is less restrictive than Bell non-locality but more restrictive than entanglement.

Although quantum discord are generally considered more general and less restrictive than entanglement, such a relationship cannot be directly observed in $\Lambda\bar{\Lambda}$ systems because entanglement and  discord all remain non-zero throughout the entire scattering angle range. Nonetheless, the remainder of the hierarchy presented in Eq. (\ref{eq:Hr}) is well-confirmed in baryon-antibaryon systems.
\subsection{Dynamical state} \label{sec:5}
The production of baryon-antibaryon pairs in $e^+e^-$ annihilation, specifically through the $J/\psi \rightarrow \Lambda \bar{\Lambda}$ decay channel, offers a high-precision platform to probe non-perturbative Quantum Chromodynamics (QCD) and the underlying hadronization mechanisms \cite{QCD1,QCD2,QCD3,QCD4,QCD5}. Within this regime, the strong interaction and quark-gluon dynamics are expected to leave distinct signatures on the spin correlations of the final-state baryons. In this study, we investigate the robustness of these quantum correlations by modeling the $\Lambda \bar{\Lambda}$ system as a bipartite qubit pair evolving through a correlated dephasing environment. 

\begin{figure}[!h]
\begin{center}
\includegraphics[width=8.5cm,height=6cm]{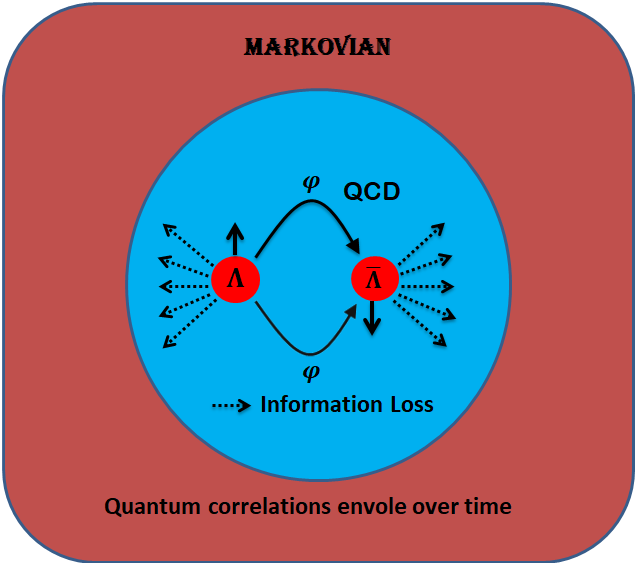}
\includegraphics[width=8.5cm,height=6cm]{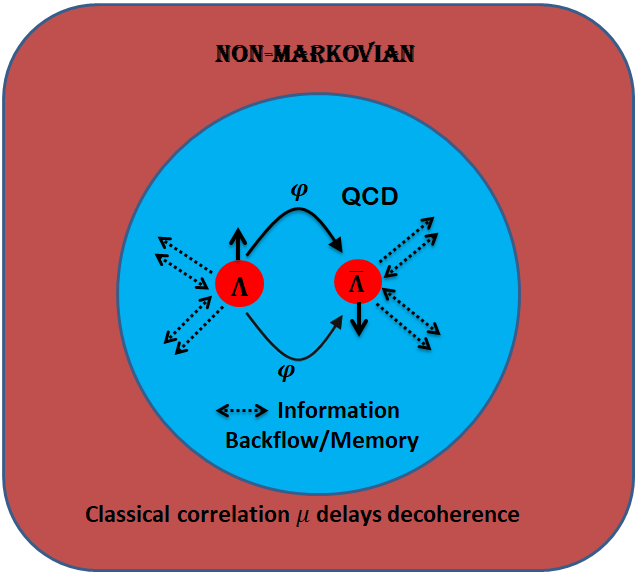}
\put(-475,145){\textcolor{white}{\fontsize{18}{20}\selectfont\bfseries (a)}}
\put(-230,145){\textcolor{white}{\fontsize{18}{20}\selectfont\bfseries (b)}}
\end{center}
\caption{Schematic illustration of the quantum evolution of $\Lambda\bar{\Lambda}$ pairs in Markovian and Non-Markovian environments.}
\label{fig:e}
\end{figure}

When two qubits traverse a channel successively, the output state is generally described by the map $\varrho(t)= \mathcal{E}[\varrho(0)]=\sum_{i,j=0}^{3}E_{i,j}\varrho(0)E_{i,j}^{\dagger}$. In the simple memoryless case, the Kraus operators are $E_{i,j}=\sqrt{p_{i}p_{j}}\sigma_i\otimes\sigma_j$, constructed from an independent probability distribution for each qubit. However, a channel may have partial memory, introducing classical correlations between successive uses. The Macchiavello-Palma model accounts for this effect via the joint probability $p_{i,j}=(1-\mu)p_i p_j + \mu p_i \delta_{i,j}$, where $\mu\in[0,1]$ measures the degree of classical correlation. Focusing on the specific case of a dephasing channel with distribution $p_0=1-p$, $p_{1,2}=0$ and $p_3=p$, one can study its time evolution using a colored noise model where the time-dependent factor is $p=\frac{1}{2}\left[1-\mathcal{V}(t)\right]$. In the non-Markovian regime (i.e. $\tau >1/4$), we have \cite{Hu}
\begin{equation}
\mathcal{V}(t)=e^{-\lambda t/2}\left[ \cos\Bigg(\frac{\sqrt{|\lambda^{2}-16|}}{2}t\Bigg)+\frac{\lambda}{\sqrt{|\lambda^{2}-16|}}\sin\Bigg(\frac{\sqrt{|\lambda^{2}-16|}}{2}t\Bigg)\right],
\end{equation}
with $\lambda=\frac{1}{\tau}$.
For the Markovian regime ($\tau <1/4$), we follow
\begin{equation}
\mathcal{V}(t)=e^{-\lambda t/2}\left[ \cosh\Bigg(\frac{\sqrt{|\lambda^{2}-16|}}{2}t\Bigg)+\frac{\lambda}{\sqrt{|\lambda^{2}-16|}}\sinh\Bigg(\frac{\sqrt{|\lambda^{2}-16|}}{2}t\Bigg)\right],
\end{equation}
Applying this formalism to a thermal input state, we obtain the time-dependent output state
\begin{equation}\label{rtt}
\varrho_{\Lambda\bar{\Lambda}}(t)=
\begin{pmatrix}
\rho_{1,1}& 0 & 0 & \mathcal{W}(t)\rho_{1,4} \\
0 & \rho_{2,2} & \mathcal{W}(t)\rho_{2,3} & 0 \\
0 & \mathcal{W}(t)\rho_{2,3} & \rho_{2,2}& 0 \\
\mathcal{W}(t)\rho_{1,4} & 0 & 0 & \rho_{4,4}
\end{pmatrix},
\end{equation}
whose off-diagonal elements are modulated by the factor $\mathcal{W}(t)= \mathcal{V}(t)^2(t) + \left[1-\mathcal{V}(t)^2(t)\right]\mu$, combining the dephasing dynamics and the degree of classical correlation.

This approach enables a systematic analysis of how environmental perturbations and channel memory effects compete with or potentially preserve the intrinsic quantum features generated during the baryon production process.

\begin{figure}[!h]
\includegraphics[scale=0.4]{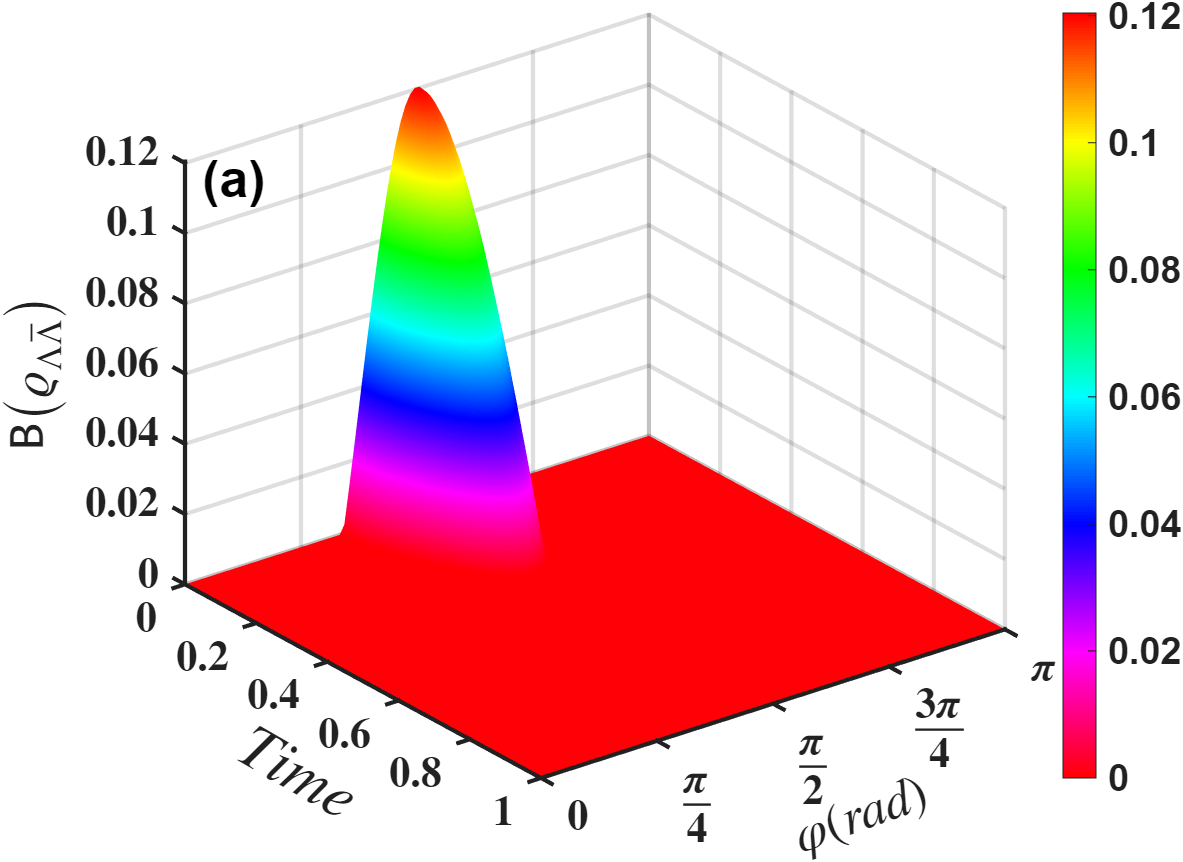}
\includegraphics[scale=0.4]{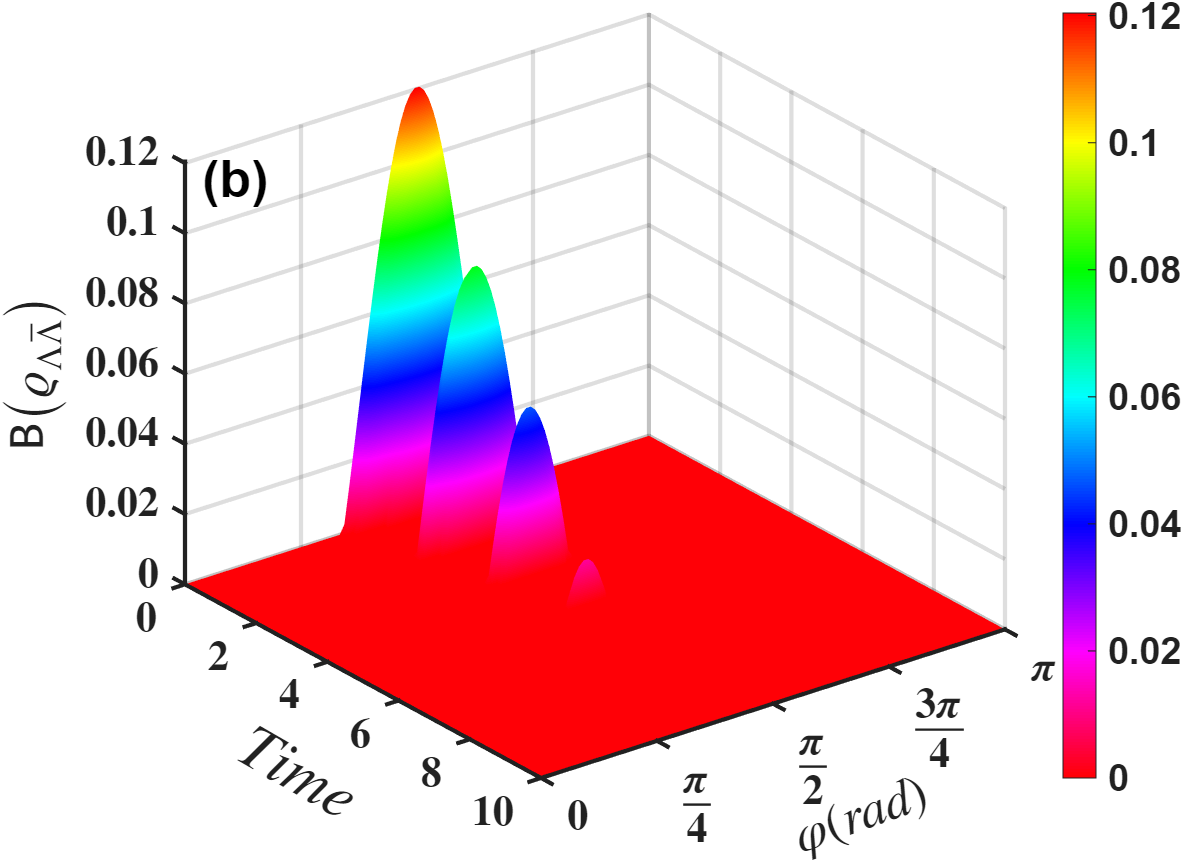}
\caption{Evolution of $\mathtt{B}\big(\varrho_{\Lambda\bar{\Lambda}}\big)$ under different environmental memory effects: (a) the Markovian case ($\tau = 0.2$) and (b) the non-Markovian case ($\tau = 20$), with mass corrections omitted and $\mu = 0.8$.}%
\label{fig:B1}
\end{figure}

\begin{figure}[!h]
\includegraphics[scale=0.4]{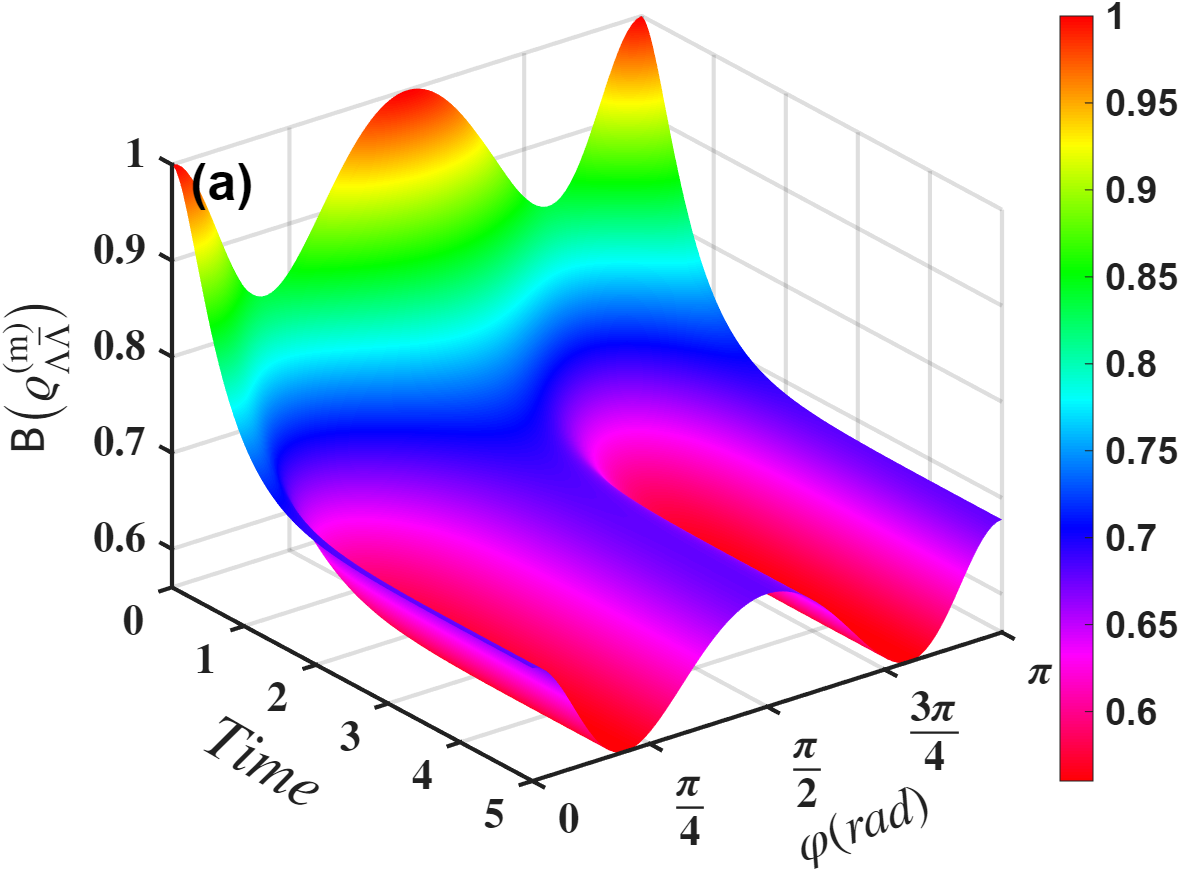}
\includegraphics[scale=0.4]{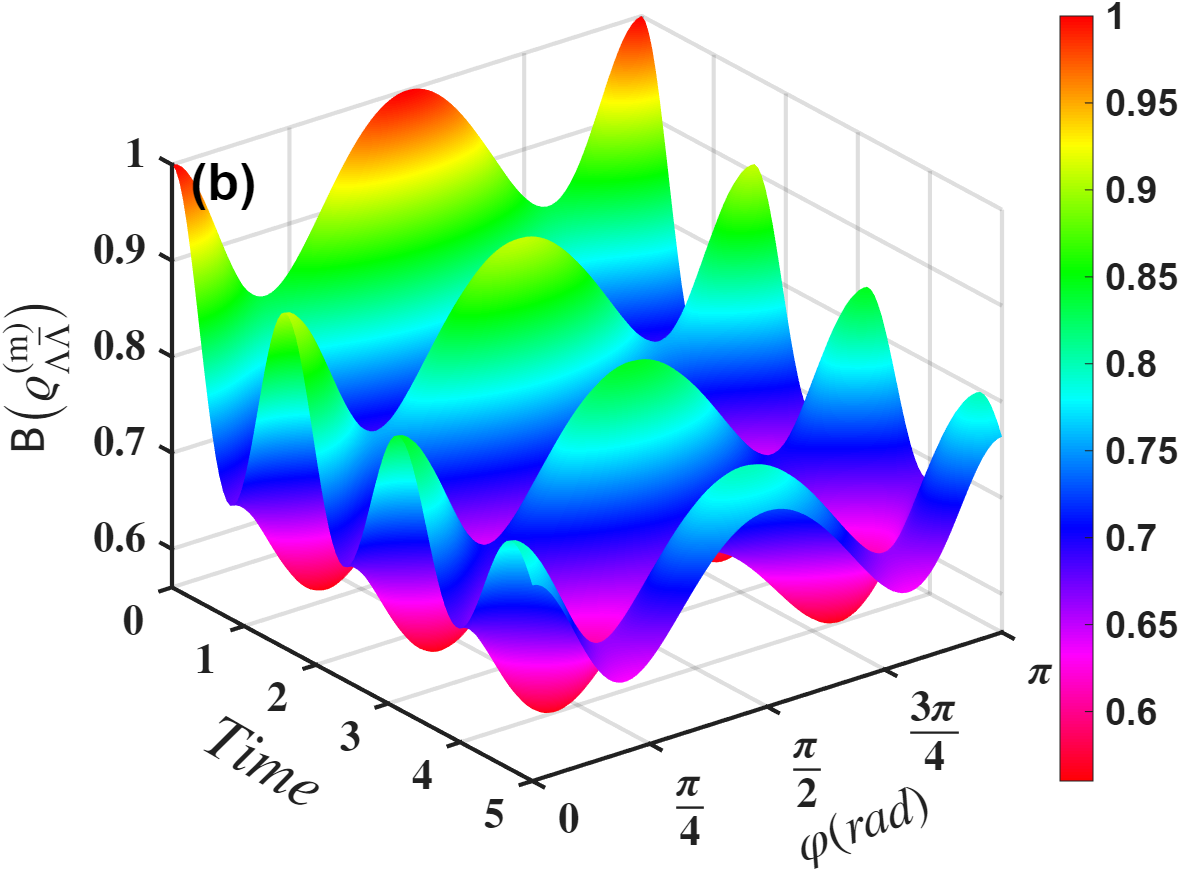}
\caption{The effect of mass corrections on the time evolution of the Bell non-locality $\mathtt{B}\big(\varrho^{\rm (m)}_{\Lambda\bar{\Lambda}}\big)$ is illustrated for $\mu = 0.8$ in two regimes: (a) Markovian ($\tau = 0.2$) and (b) non-Markovian ($\tau = 5$).}
\label{fig:B2}
\end{figure} 

We have plotted Bell nonlocality under different conditions to analyze the effects of the scattering angle $\varphi$, time evolution, and mass-dependent corrections. It is observed that the Bell violation reaches its maximum at $\varphi = \pi/2$ and decreases monotonically as time increases or as $\varphi$ moves away from $\pi/2$. Furthermore, Bell non-locality is a function symmetric with respect to $\varphi = \pi/2$ over the interval $[0, \pi]$, and its maximum value is attained precisely at $\varphi = \pi/2$. The inclusion of mass-dependent corrections does not modify the qualitative angular symmetry, but it quantitatively alters the maximal value of the Bell parameter and slightly reshapes the structure of the extrema.

Fig. \ref{fig:B1}(a) presents the Markovian dynamics of Bell non-locality $\mathtt{B}\big(\varrho_{\Lambda\bar{\Lambda}}\big)$ as a function of the scattering angle $\varphi$. For $\varphi=\pi/2$, this leads to higher initial values of the Bell parameter. The Bell parameter then decreases, reflecting a reduction in non-classical correlations. Therefore, selecting $\varphi$ within a specific range is crucial for achieving both a high initial $\mathtt{B}_{\text{max}}$ and long-term stability of non-classical correlations. In the non-Markovian regime [see Fig. \ref{fig:B1}(b)], we observe a consistent trend, though it is characteristically modulated by the presence of memory effects.

When mass-dependent corrections are included, the angular structure is modified quantitatively. In addition to the dominant maximum at $\varphi=\pi/2$, secondary maxima also appear at $\varphi=0$, and $\varphi=\pi$, as observed in Fig.\ref{fig:B2}(a). Although the qualitative temporal decay remains unchanged, the inclusion of mass effects reshapes the angular distribution of the extrema and alters the maximal attainable value of the Bell parameter. This behavior is also observed in the non-Markovian regime [see Fig. \ref{fig:B2}(b)], where memory effects modulate the temporal evolution but preserve the underlying angular structure induced by both the dominant dynamical term and its mass-dependent corrections.

\begin{figure}[!h]
\includegraphics[scale=0.4]{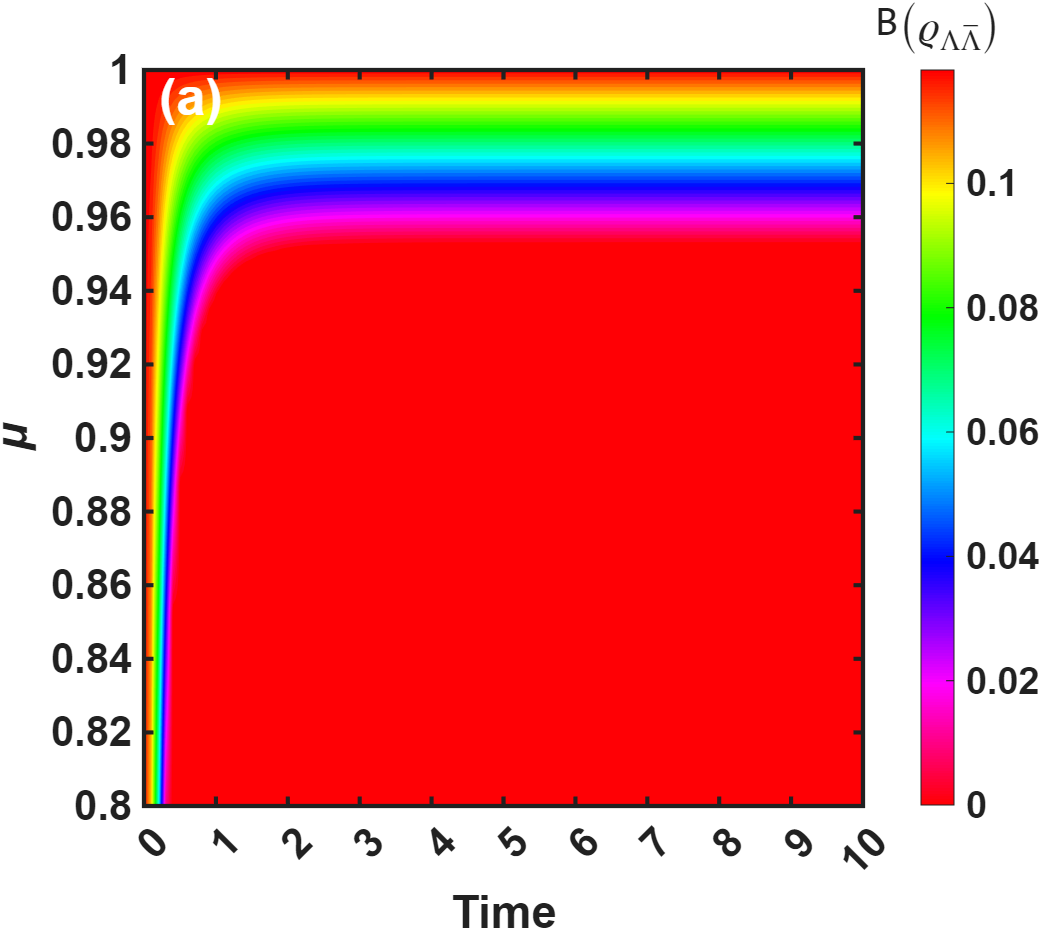}
\includegraphics[scale=0.4]{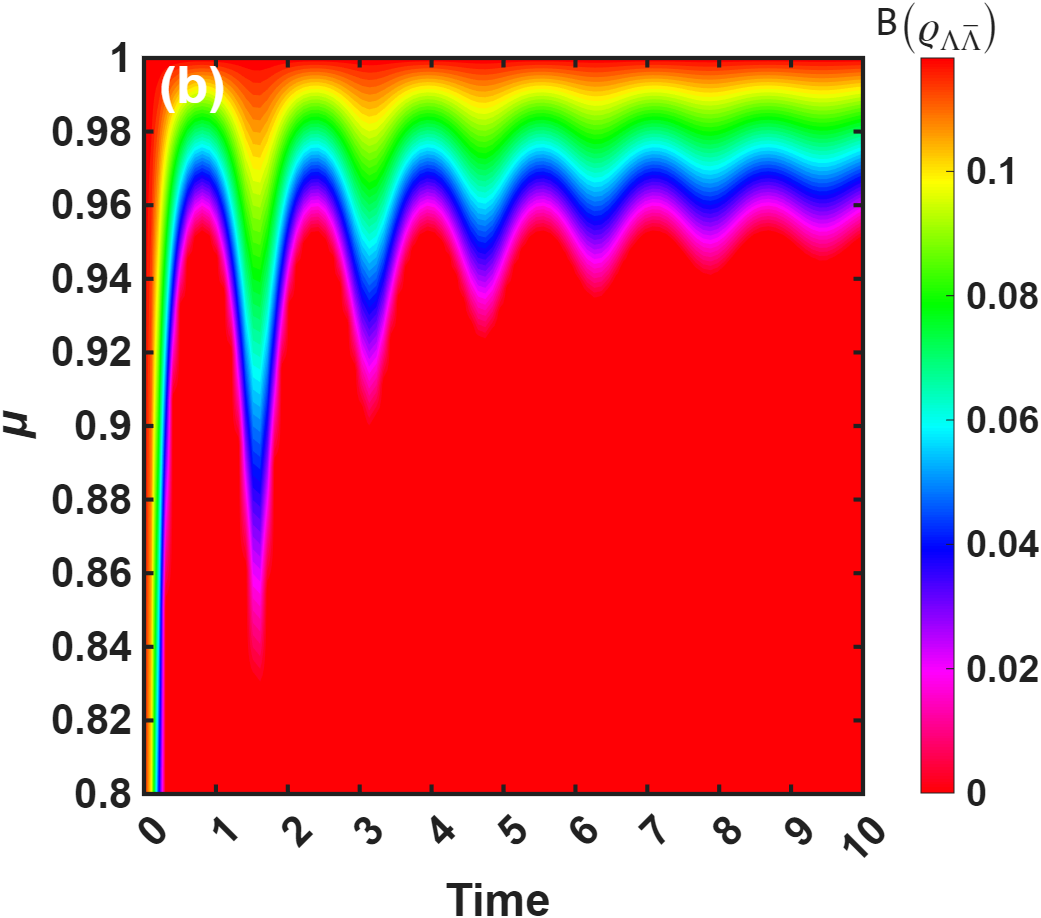}
\caption{Evolution of $\mathtt{B}\big(\varrho_{\Lambda\bar{\Lambda}}\big)$ under different environmental memory effects: (a) the Markovian case ($\tau = 0.2$) and (b) the non-Markovian case ($\tau = 20$), with mass corrections omitted and $\varphi = \pi/2$.}
\label{fig:Mu1}
\end{figure}
\begin{figure}[!h]
\includegraphics[scale=0.4]{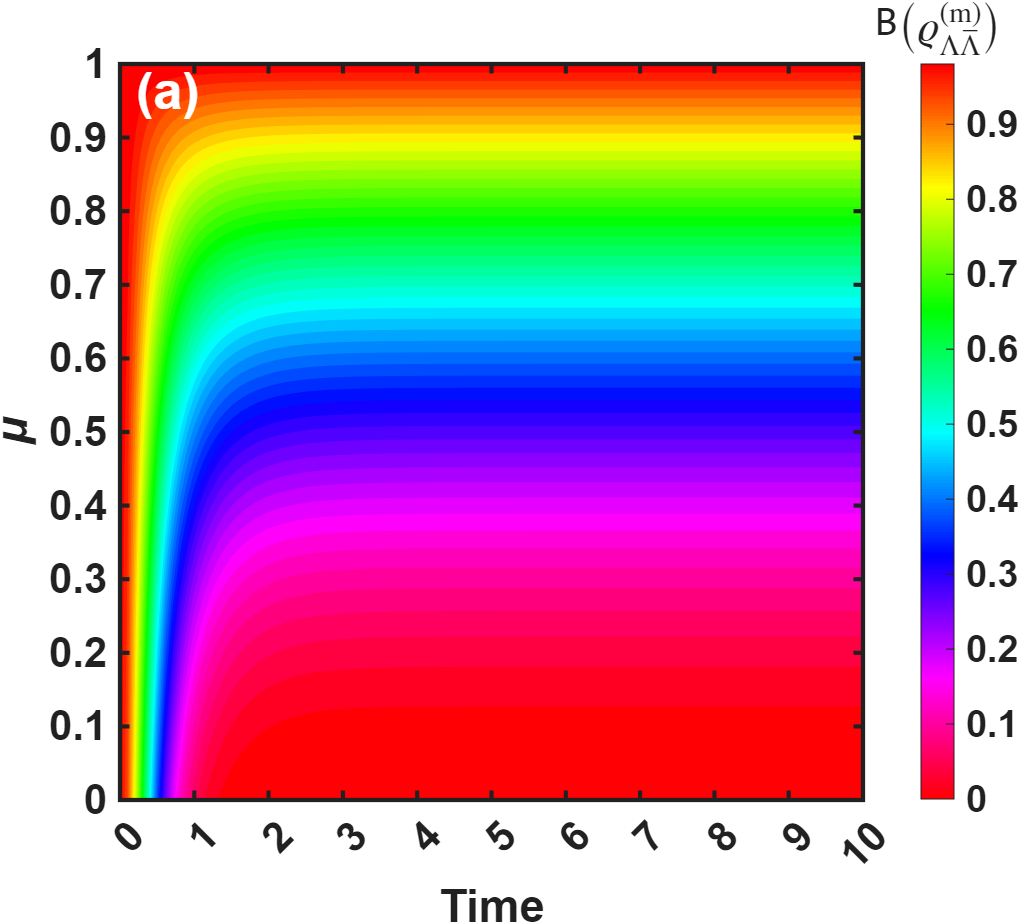}
\includegraphics[scale=0.4]{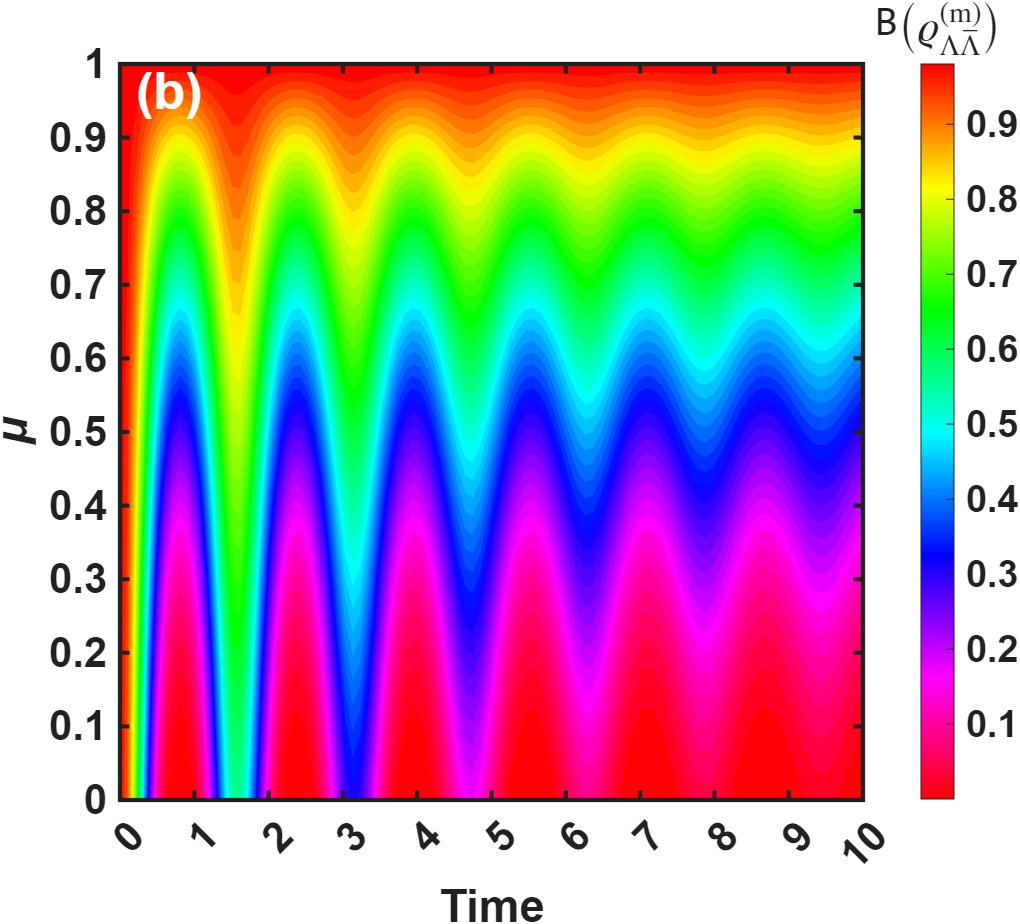}
\caption{The effect of mass corrections on the time evolution of the Bell non-locality $\mathtt{B}\big(\varrho^{\rm (m)}_{\Lambda\bar{\Lambda}}\big)$ is illustrated for $\varphi=\pi/2$ in (a) the Markovian regime ($\tau = 0.2$) and (b) the non-Markovian regime ($\tau = 5$).}
\label{fig:Mu2}
\end{figure}

The impact of the classical parameter $\mu$ on the evolution of the Bell parameter is depicted in Figs. \ref{fig:Mu1} and \ref{fig:Mu2}. We find that stronger correlations provide a buffer against the rapid decay of $\mathtt{B}_{\text{max}}$. In the non-Markovian regime, this protective effect persists, though the resulting dynamics are shaped by memory-induced, non-exponential relaxation processes. As $\mu$ approaches unity, the Bell parameter exhibits minimal decay, indicating that non-classical correlations remain remarkably stable throughout the system's evolution.

A comparison between the mass-independent and mass-corrected cases reveals that the inclusion of mass-dependent corrections enhances the maximal violation of Bell inequality. In particular, the Bell parameter reaches a higher maximum value in the presence of mass corrections than in the massless case. While the qualitative temporal behavior remains unchanged, mass effects quantitatively modify the angular distribution and introduce additional extrema at $\varphi=0$ and $\varphi=\pi/2$. This indicates that mass corrections reinforce the strength of nonlocal correlations without altering their fundamental dynamical origin.

\begin{figure}[!h]
\includegraphics[scale=0.4]{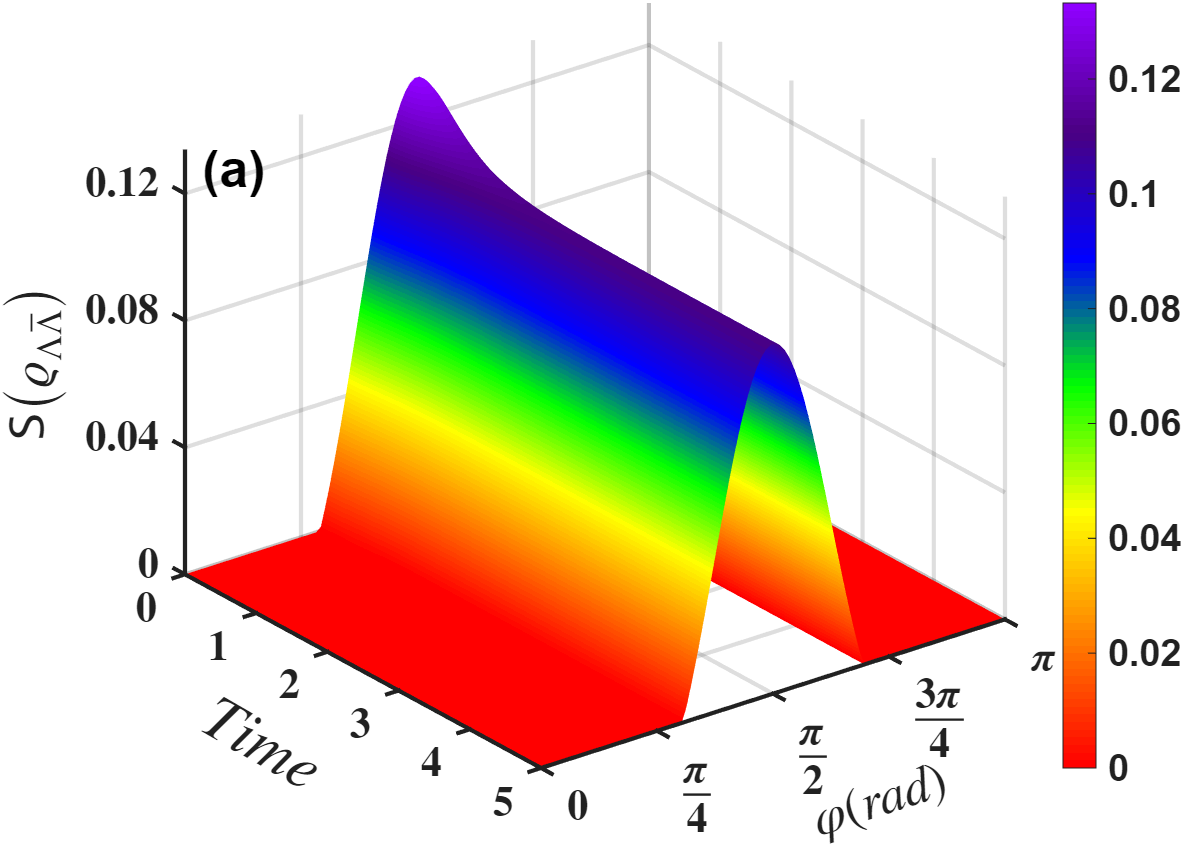}
\includegraphics[scale=0.4]{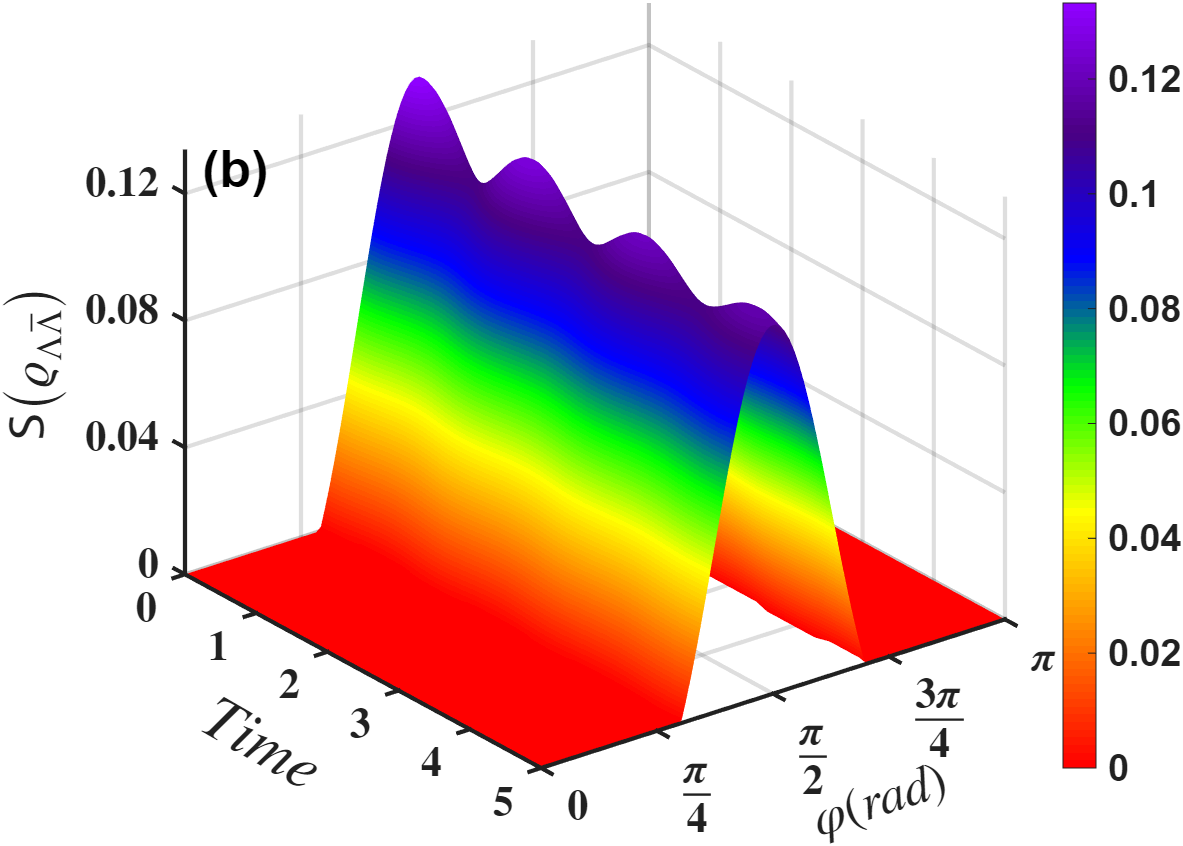}
\caption{Evolution of $\mathtt{S}\big(\varrho_{\Lambda\bar{\Lambda}}\big)$ under different environmental memory effects: (a) the Markovian case ($\tau = 0.2$) and (b) the non-Markovian case ($\tau = 5$), with mass corrections omitted and $\mu = 0.8$.}
\label{fig:ch1}
\end{figure}
\begin{figure}[!h]
\includegraphics[scale=0.4]{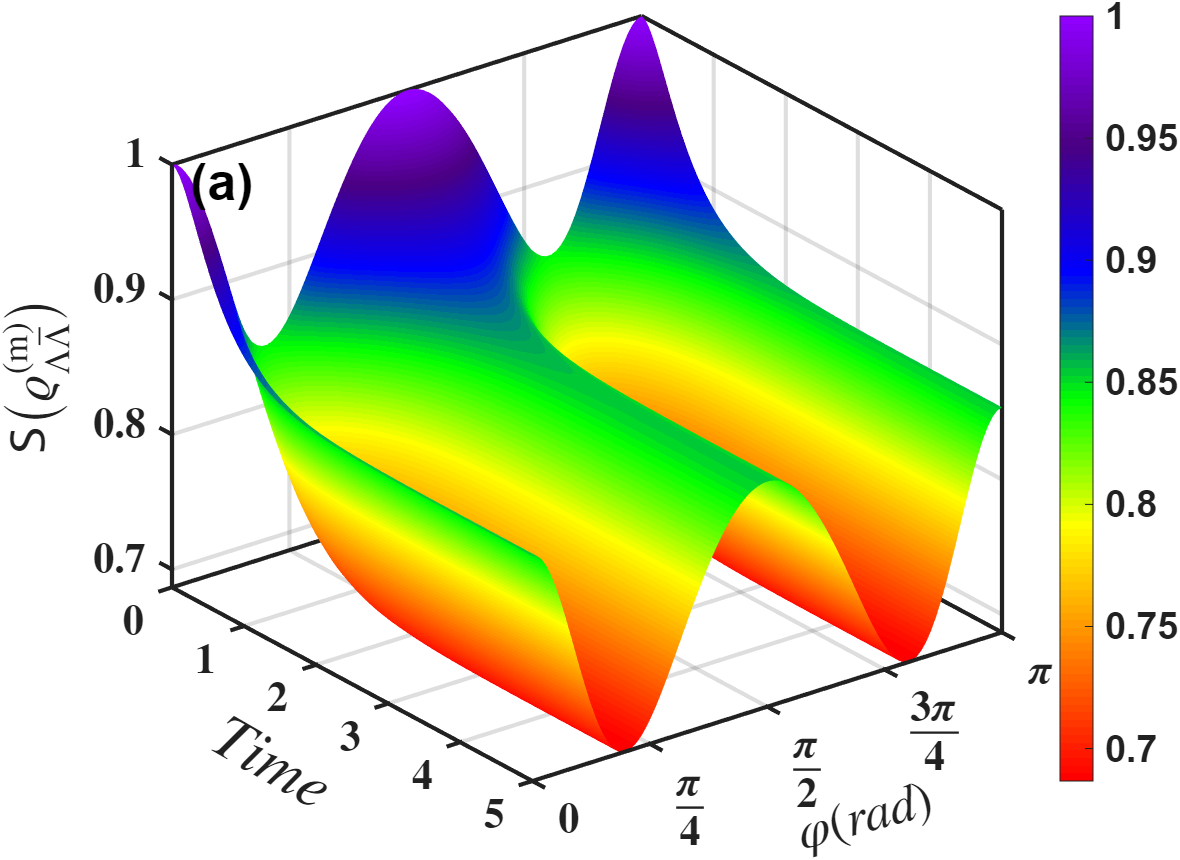}
\includegraphics[scale=0.4]{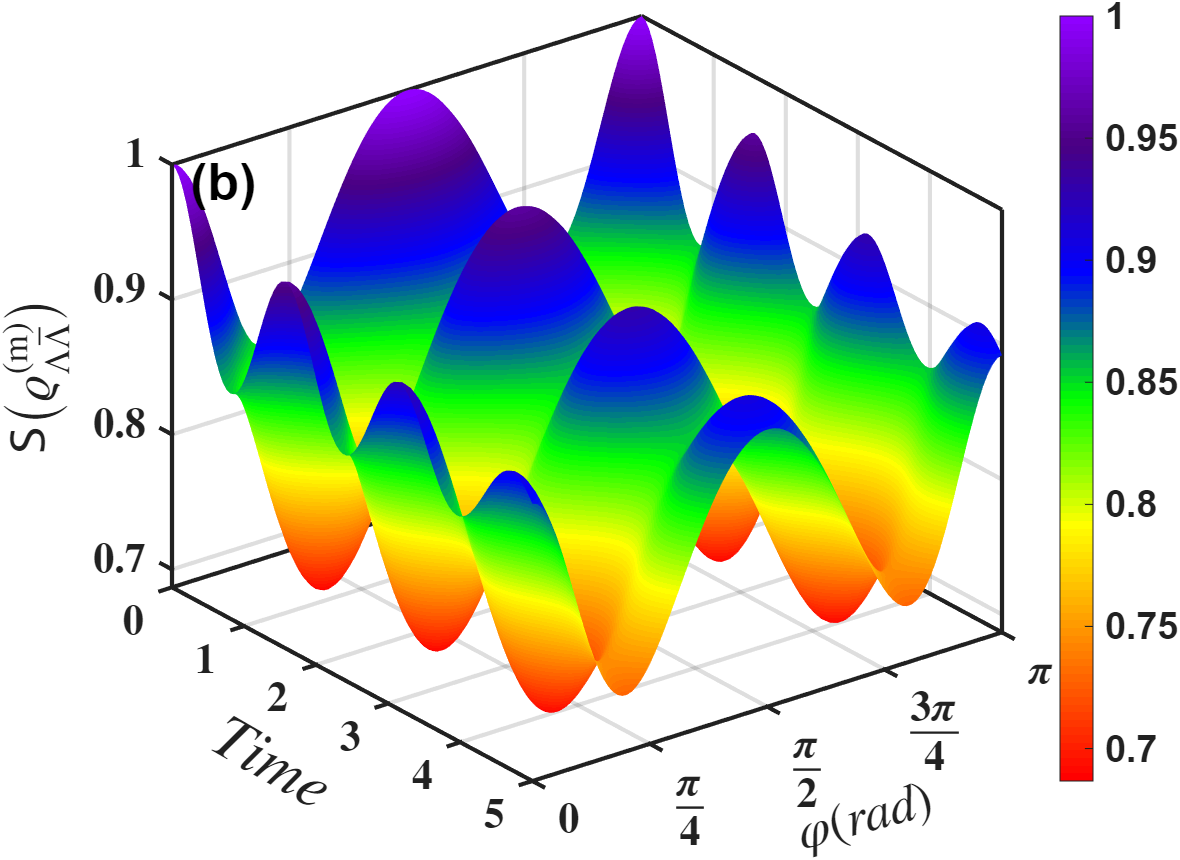}
\caption{The effect of mass corrections on the time evolution of the quantum steering $\mathtt{S}\big(\varrho^{\rm (m)}_{\Lambda\bar{\Lambda}}\big)$ is illustrated for $\mu = 0.8$ in (a) the Markovian regime ($\tau = 0.2$) and (b) the non-Markovian regime ($\tau = 5$).}
\label{fig:ch2}
\end{figure}

For identical time intervals, the evolution of quantum steering is illustrated in Fig.~\ref{fig:ch1}(a) for the Markovian regime and in Fig.~\ref{fig:ch1}(b) for the non-Markovian case, considering the mass-independent scenario. In the Markovian limit, the steering $\mathtt{S}(\varrho_{\Lambda\bar{\Lambda}})$ exhibits a gradual decay with time, reflecting the irreversible loss of quantum correlations induced by the environment. As shown in Fig.~\ref{fig:ch1}(a), the maximal steerability occurs near the scattering angle $\varphi=\pi/2$, indicating that the relative spin orientation of the produced baryon–antibaryon pair plays a crucial role in enhancing quantum correlations. In addition, the steering distribution displays a clear symmetry around $\varphi=\pi/2$.

A markedly different behavior emerges in the non-Markovian regime, as depicted in Fig.~\ref{fig:ch1}(b). Instead of a monotonic decay, the steerability exhibits oscillatory dynamics characterized by temporal revivals. These oscillations originate from memory effects of the environment, which induce a backflow of information from the environment to the $\Lambda\bar{\Lambda}$ system and temporarily restore the lost quantum correlations. Nevertheless, the largest steering values remain concentrated around $\varphi=\pi/2$.

The impact of mass corrections is illustrated in Fig.~\ref{fig:ch2}(a), where the evolution of $\mathtt{S}(\varrho^{\rm (m)}_{\Lambda\bar{\Lambda}})$ is shown within the Markovian regime. The results reveal that including mass-dependent contributions leads only to minor quantitative modifications, leaving both the maximal steering values and the global angular dependence essentially unchanged.

The non-Markovian dynamics with mass corrections, presented in Fig.~\ref{fig:ch2}(b), display pronounced oscillatory features and partial revivals of steering. These effects reflect the non-Markovian information backflow that intermittently restores quantum correlations in the $\Lambda\bar{\Lambda}$ pair. Although the maximal steerability continues to occur near $\varphi=0$, $\varphi=\pi/2$, and $\varphi=\pi$, the presence of memory effects generates a richer dynamical structure characterized by periodic restorations of quantum correlations.

\begin{figure}[!h]
\includegraphics[scale=0.4]{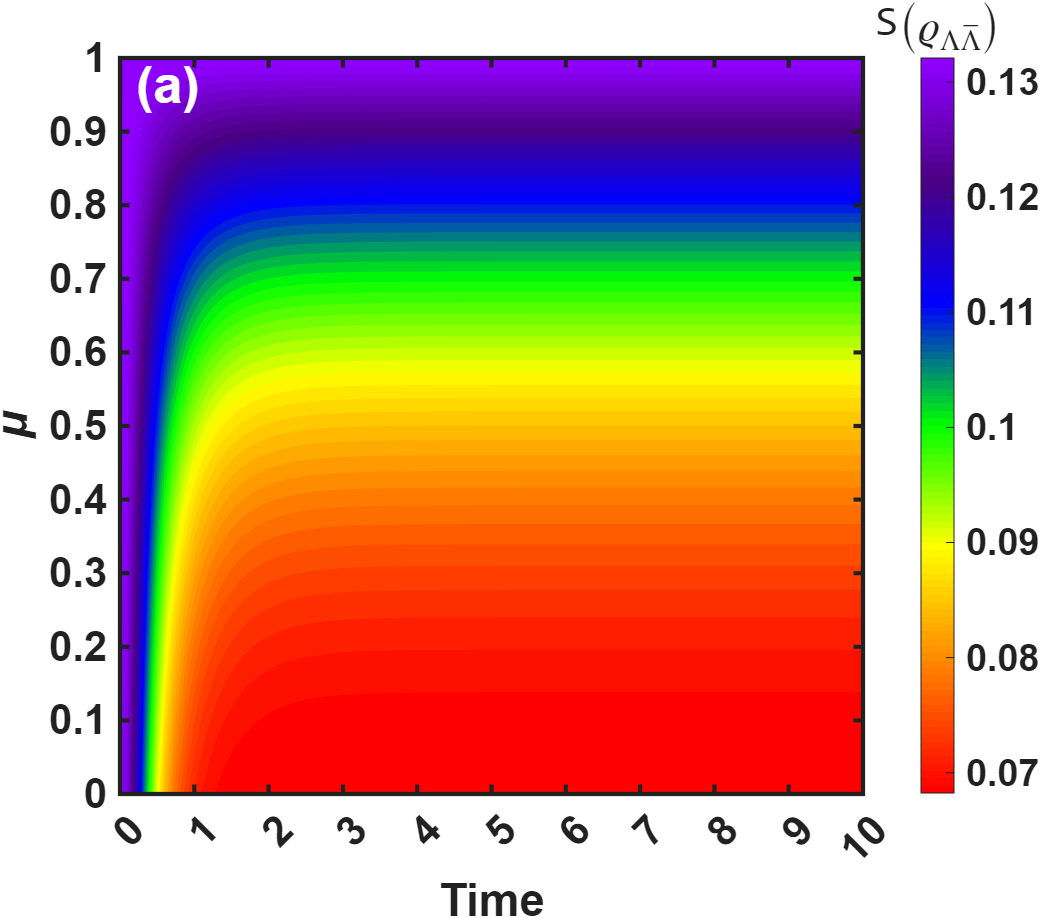}
\includegraphics[scale=0.4]{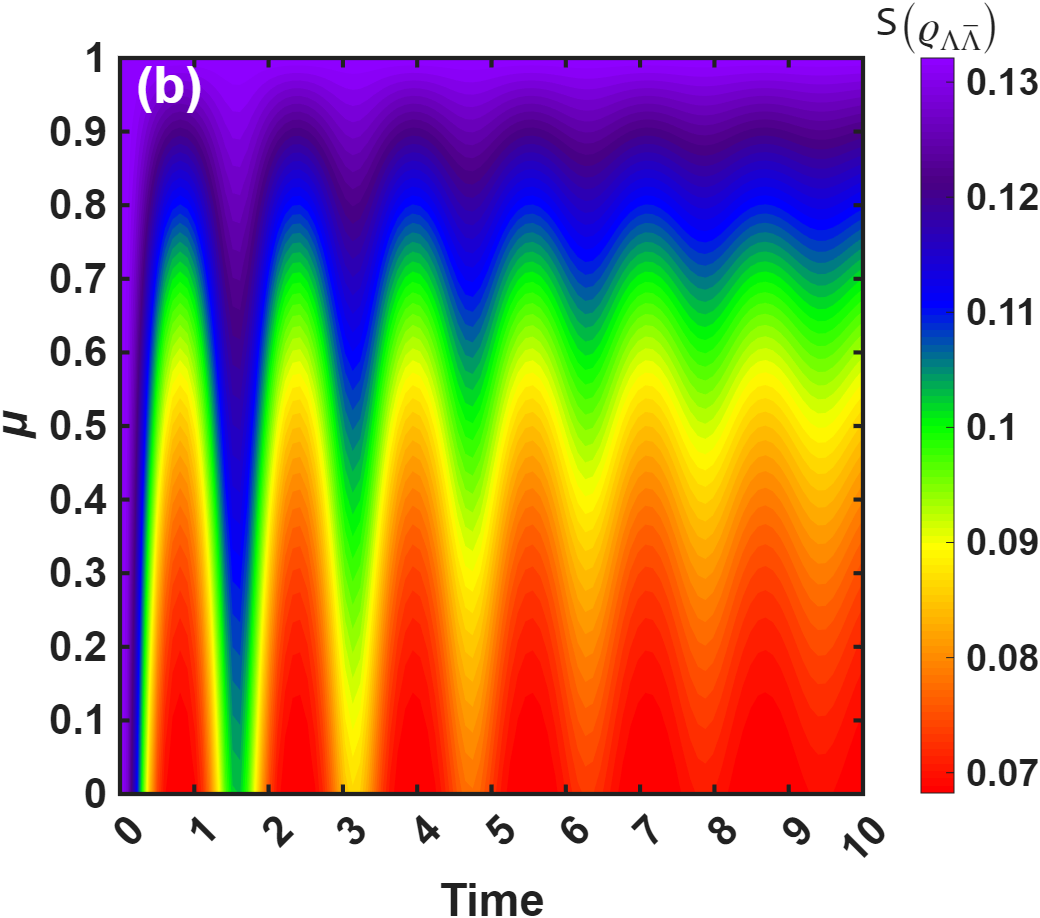}
\caption{Evolution of $\mathtt{S}\big(\varrho_{\Lambda\bar{\Lambda}}\big)$ under different environmental memory effects: (a) the Markovian case ($\tau = 0.2$) and (b) the non-Markovian case ($\tau = 5$), with mass corrections omitted and $\varphi = \pi/2$.}
\label{fig:ch3}
\end{figure}
\begin{figure}[!h]
\includegraphics[scale=0.4]{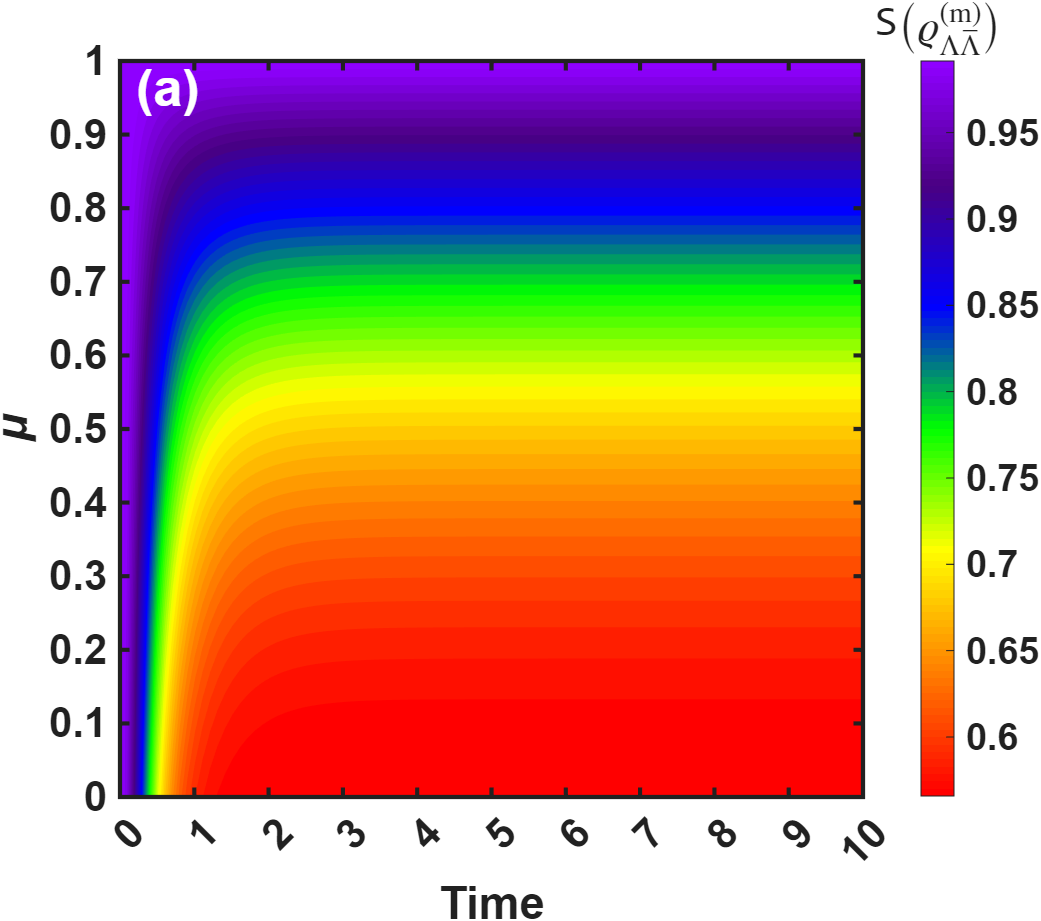}
\includegraphics[scale=0.4]{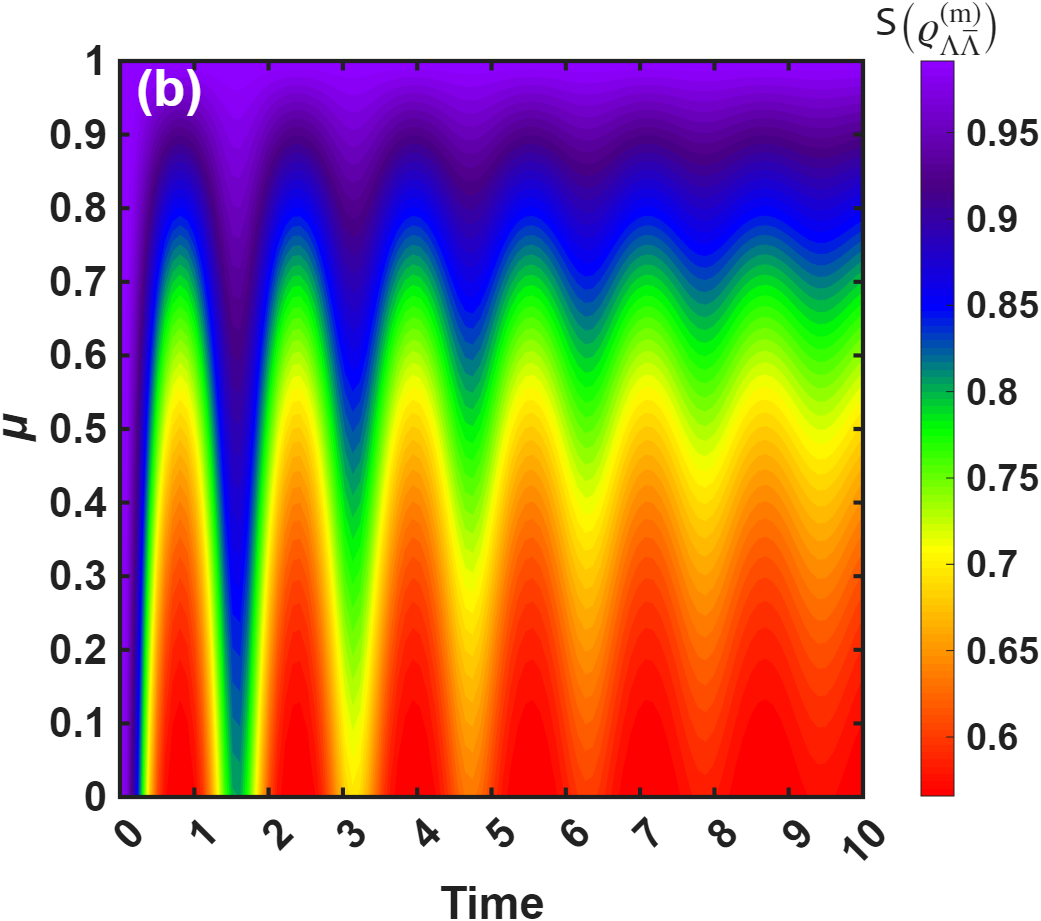}
\caption{The effect of mass corrections on the time evolution of the quantum steering $\mathtt{S}\big(\varrho^{\rm (m)}_{\Lambda\bar{\Lambda}}\big)$ is illustrated for $\varphi=\pi/2$ in (a) the Markovian regime ($\tau = 0.2$) and (b) the non-Markovian regime ($\tau = 5$).}
\label{fig:ch4}
\end{figure}

The evolution of steerability is strongly influenced by the degree of classical correlations, controlled by the parameter $\mu$. Figure~\ref{fig:ch3} presents the behavior of $\mathtt{S}\big(\varrho_{\Lambda\bar{\Lambda}}\big)$ as a function of time and $\mu$. As the system evolves, the steerability gradually increases and eventually saturates at a constant value, indicating the establishment of a steady-state regime. Furthermore, the decay rate of steerability decreases as $\mu$ increases, revealing that stronger classical correlations in the environment effectively mitigate the decoherence process and protect the quantum correlations of the system.

The influence of correlated noise is further illustrated in Fig.~\ref{fig:ch4}, where the oscillatory dynamics of quantum steering are shown for different values of $\mu$. As the classical correlation parameter grows, the amplitude of the oscillations becomes progressively smaller, while the decay of steerability occurs more slowly. This behavior highlights the protective role of classical correlations, which reduce the detrimental impact of environmental noise and contribute to the preservation of steerability.

We investigate the temporal and angular evolution of the quantum steering for $\Lambda\bar{\Lambda}$ pairs produced in baryon decays, considering both Markovian and non-Markovian environments. Our results demonstrate a fundamental divergence in the decoherence pathways: while the Markovian limit is characterized by a monotonic and irreversible decay of quantum correlations, the non-Markovian regime exhibits distinct oscillatory patterns and temporal revivals. These oscillations are attributed to memory effects and a backflow of information from the environment to the $\Lambda\bar{\Lambda}$ system. We show that while mass corrections have a negligible impact on the peak steerability values, the alignment of particle spins at specific angles ($\varphi=\pi/2$) remains a critical factor for maximizing quantum correlations. Furthermore, we analyze the influence of the classical correlation parameter $\mu$, finding that stronger classical correlations significantly delay the decoherence rate and suppress the amplitude of non-Markovian oscillations. These findings suggest that classical correlations act as a stabilizing mechanism for quantumness in bipartite baryon systems, providing insights into the preservation of entanglement in open quantum systems at high energies.

\begin{figure}[!h]
\includegraphics[scale=0.4]{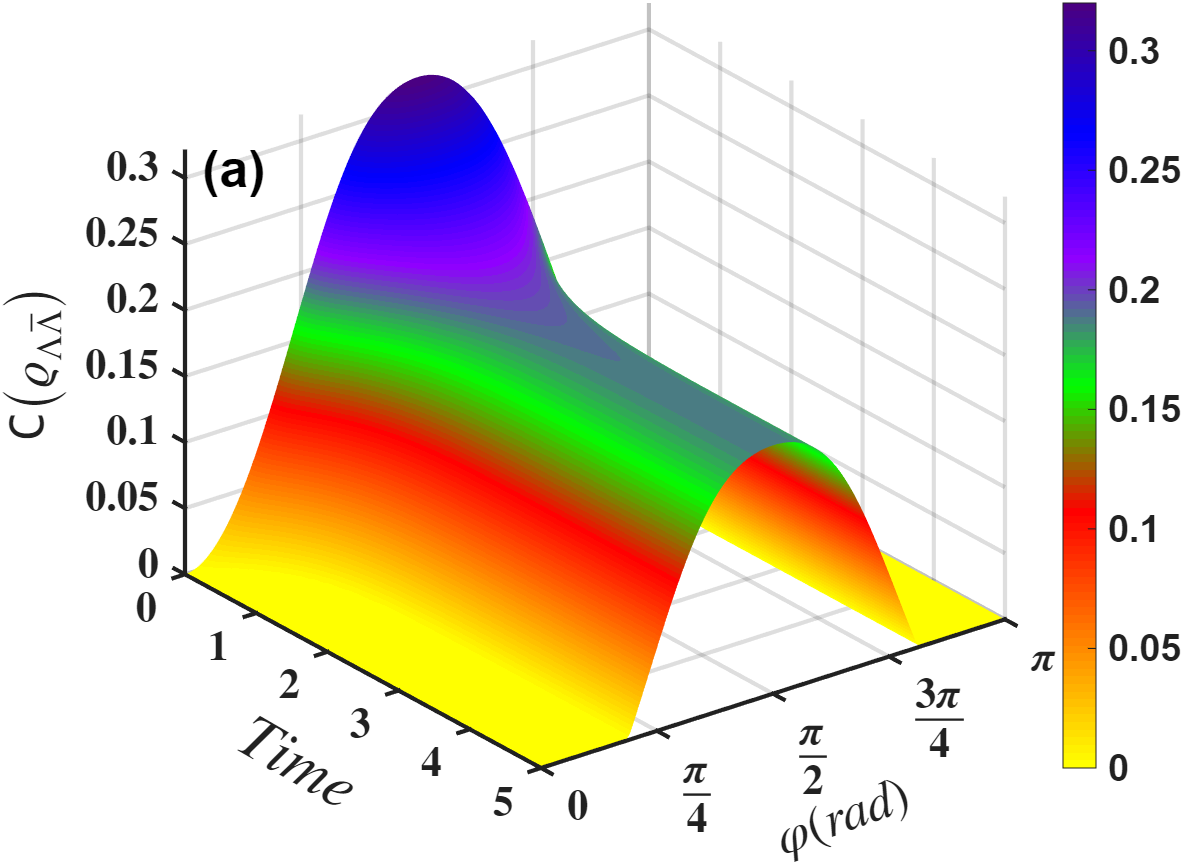}
\includegraphics[scale=0.4]{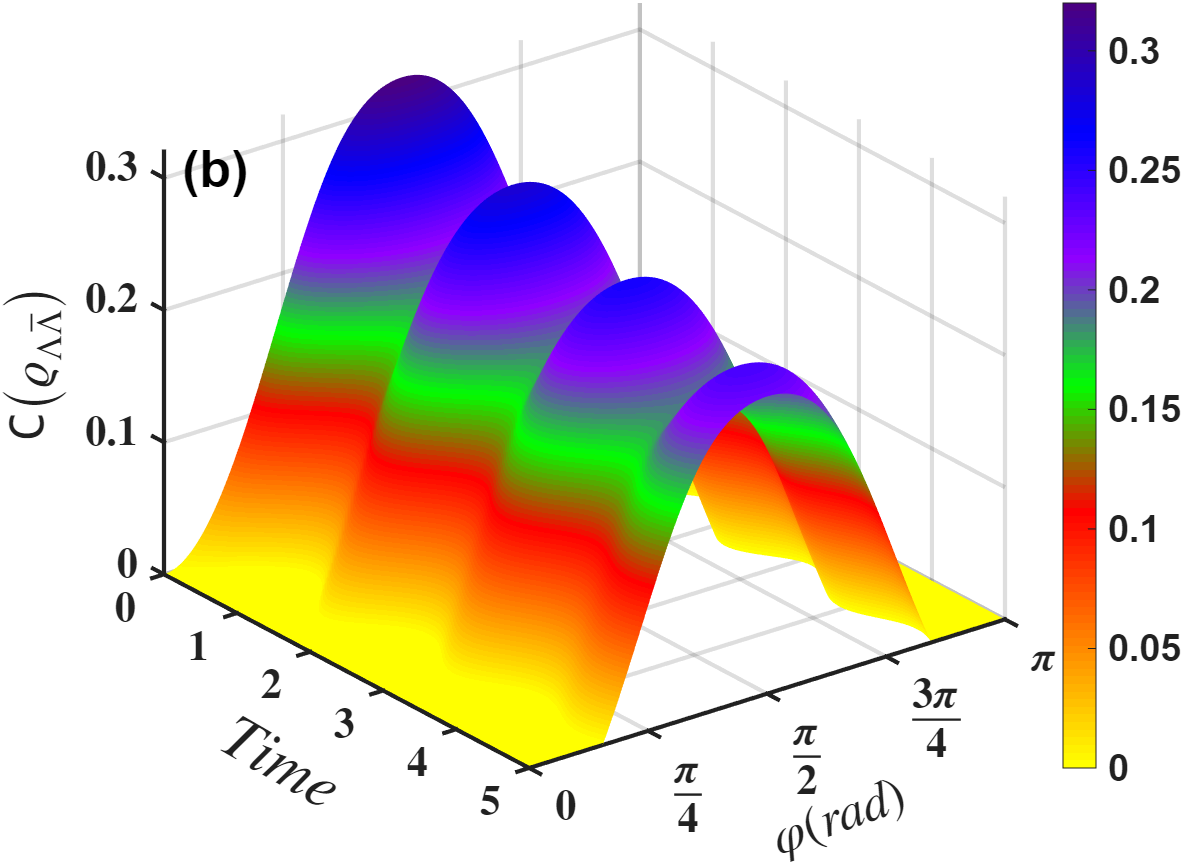}
\caption{Evolution of $\mathtt{C}\big(\varrho_{\Lambda\bar{\Lambda}}\big)$ under different environmental memory effects: (a) the Markovian case ($\tau = 0.2$) and (b) the non-Markovian case ($\tau = 5$), with mass corrections omitted and $\mu = 0.8$.}
\label{fig:e1}
\end{figure}
\begin{figure}[!h]
\includegraphics[scale=0.4]{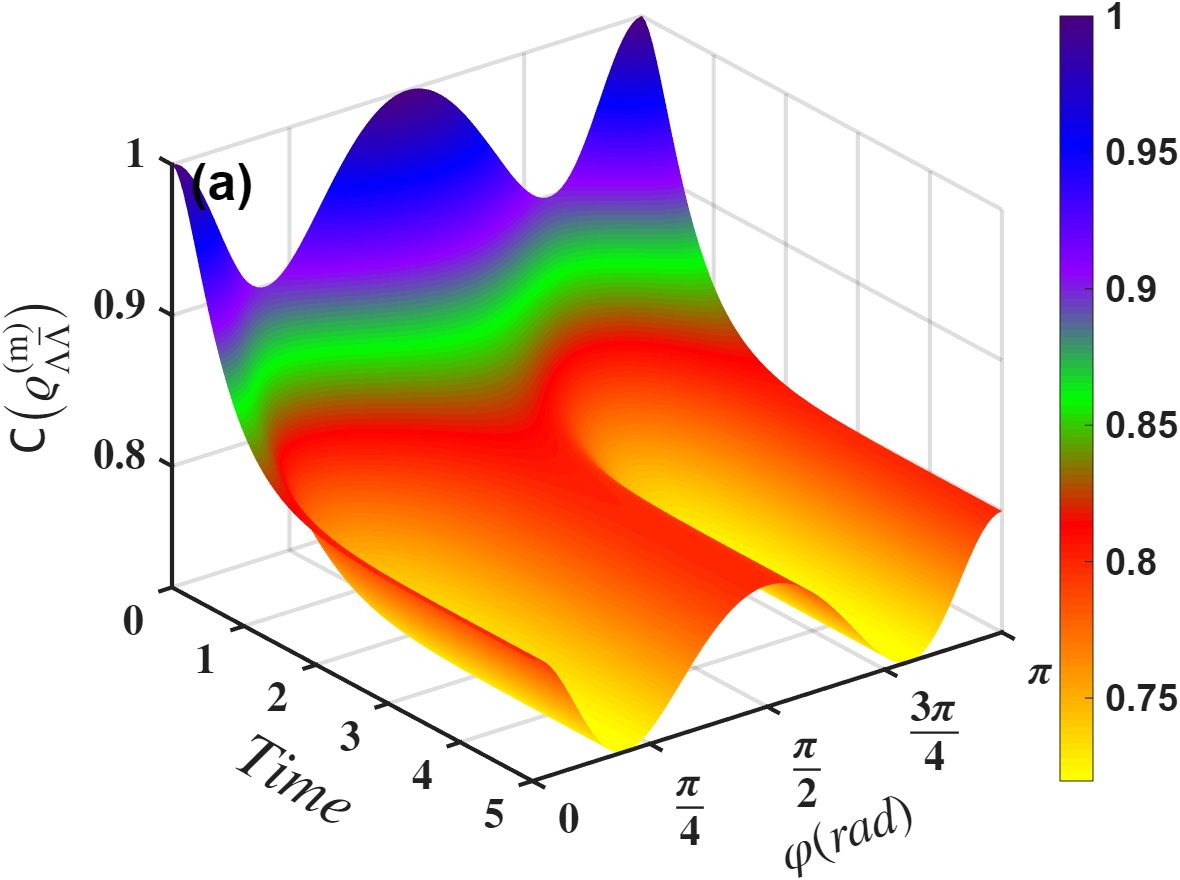}
\includegraphics[scale=0.4]{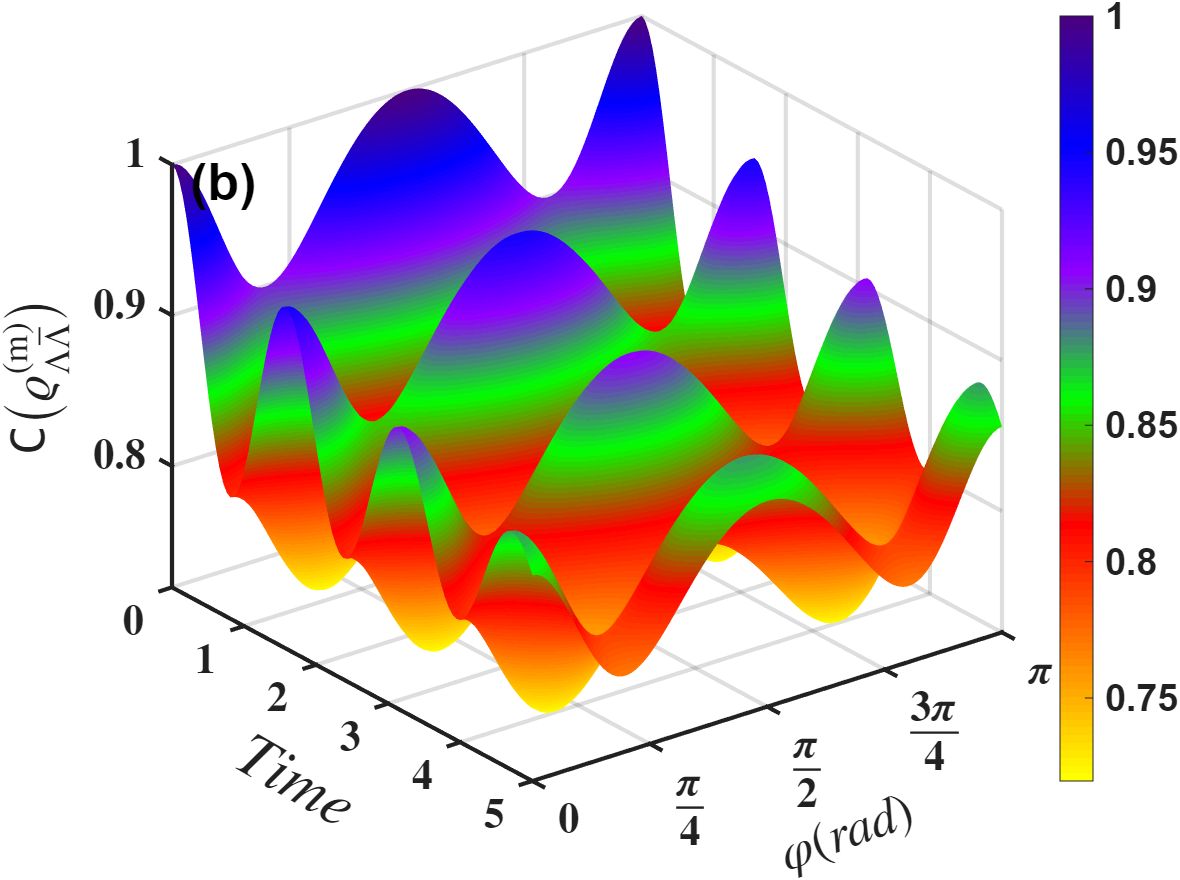}
\caption{The effect of mass corrections on the time evolution of the concurrence $\mathtt{C}\big(\varrho^{\rm (m)}_{\Lambda\bar{\Lambda}}\big)$ is illustrated for $\mu = 0.8$ in (a) the Markovian regime ($\tau = 0.2$) and (b) the non-Markovian regime ($\tau = 5$).}
\label{fig:e2}
\end{figure}
Figure~\ref{fig:e1}(a) displays the time and angular evolution of the concurrence $\mathtt{C}\big(\varrho_{\Lambda\bar{\Lambda}}\big)$ in the Markovian regime. As the system evolves, the entanglement gradually decreases due to the dissipative interaction with the environment. The maximal concurrence occurs near the scattering angle $\varphi=\pi/2$, indicating that this configuration optimally enhances quantum correlations between the baryon and antibaryon. Moreover, the angular dependence of the concurrence exhibits a pronounced symmetry around $\varphi=\pi/2$, while the entanglement decreases as $\varphi$ deviates from this value.

A qualitatively different behavior appears in the non-Markovian regime, as illustrated in Fig.~\ref{fig:e1}(b). In this case, the concurrence $\mathtt{C}\big(\varrho_{\Lambda\bar{\Lambda}}\big)$ no longer shows a monotonic decay. Instead, it undergoes oscillatory dynamics with partial revivals of entanglement. These oscillations originate from environmental memory effects, which induce a temporary backflow of information from the environment to the $\Lambda\bar{\Lambda}$ system and partially restore the degraded quantum correlations. Despite this richer dynamical behavior, the maximal entanglement remains concentrated around $\varphi=\pi/2$.

As shown in Fig. \ref{fig:e2}(a), we plot the evolution of the concurrence $\mathtt{C}\big(\varrho^{\rm (m)}_{\Lambda\bar{\Lambda}}\big)  $ as a function of time and the scattering angle $  \varphi  $ in the Markovian regime. When mass corrections are taken into account, the maximum value of the entanglement is significantly enhanced. These same mass effects quantitatively alter the angular dependence on $  \varphi  $ and cause the appearance of additional extrema at $  \varphi=0$ and $  \varphi=\pi$.

We have plotted the same function in the non-Markovian regime. The analysis is analogous to that performed in the Markovian case, as shown in \ref{fig:e2}(b). However, due to memory effects, the dynamics of concurrence no longer exhibits a monotonic decay over time. Instead, it shows characteristic oscillations, reflecting an exchange of information between the system and its environment. The maximum value of entanglement is still reached around $\varphi = \pi/2$, and the mass corrections continue to quantitatively modify the angular distribution by introducing additional extrema, particularly at $\varphi = 0$ and $\varphi = \pi$. Nevertheless, the presence of non-Markovianity leads to a richer dynamical behavior, marked by temporary revivals of entanglement.  

\begin{figure}[!h]
\includegraphics[scale=0.4]{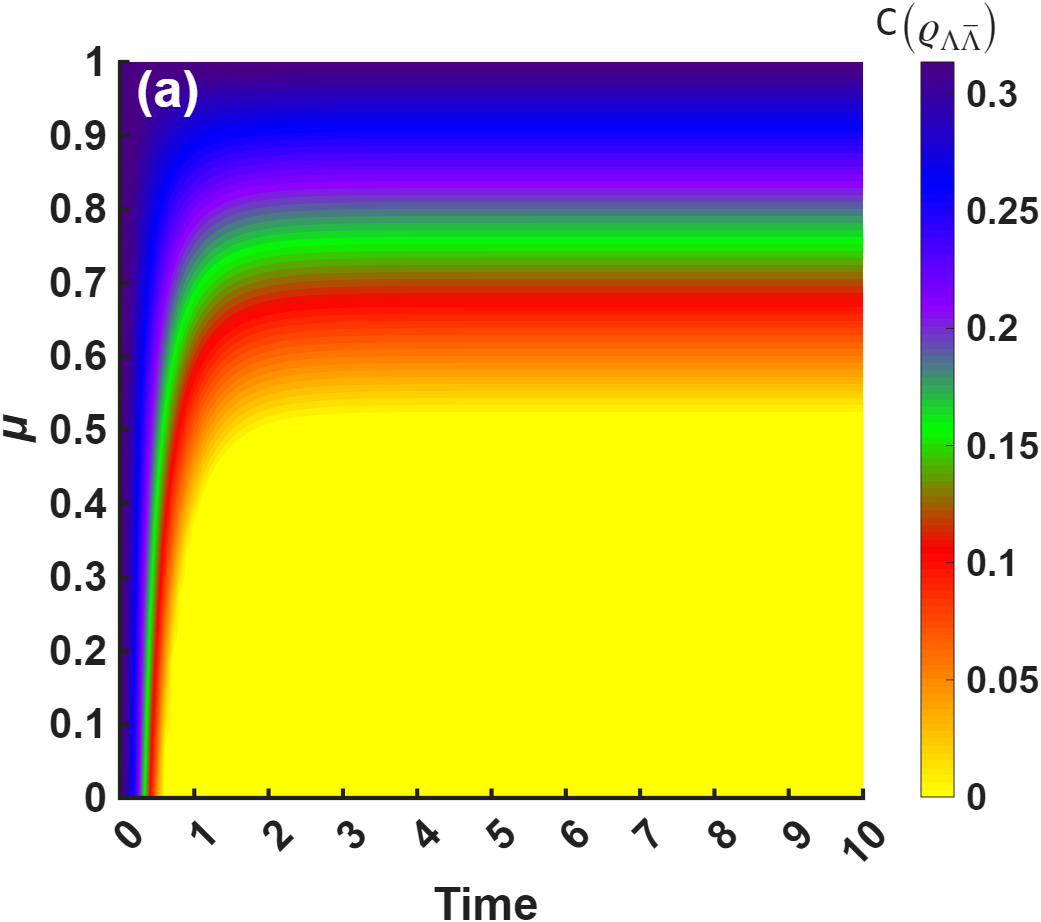}
\includegraphics[scale=0.4]{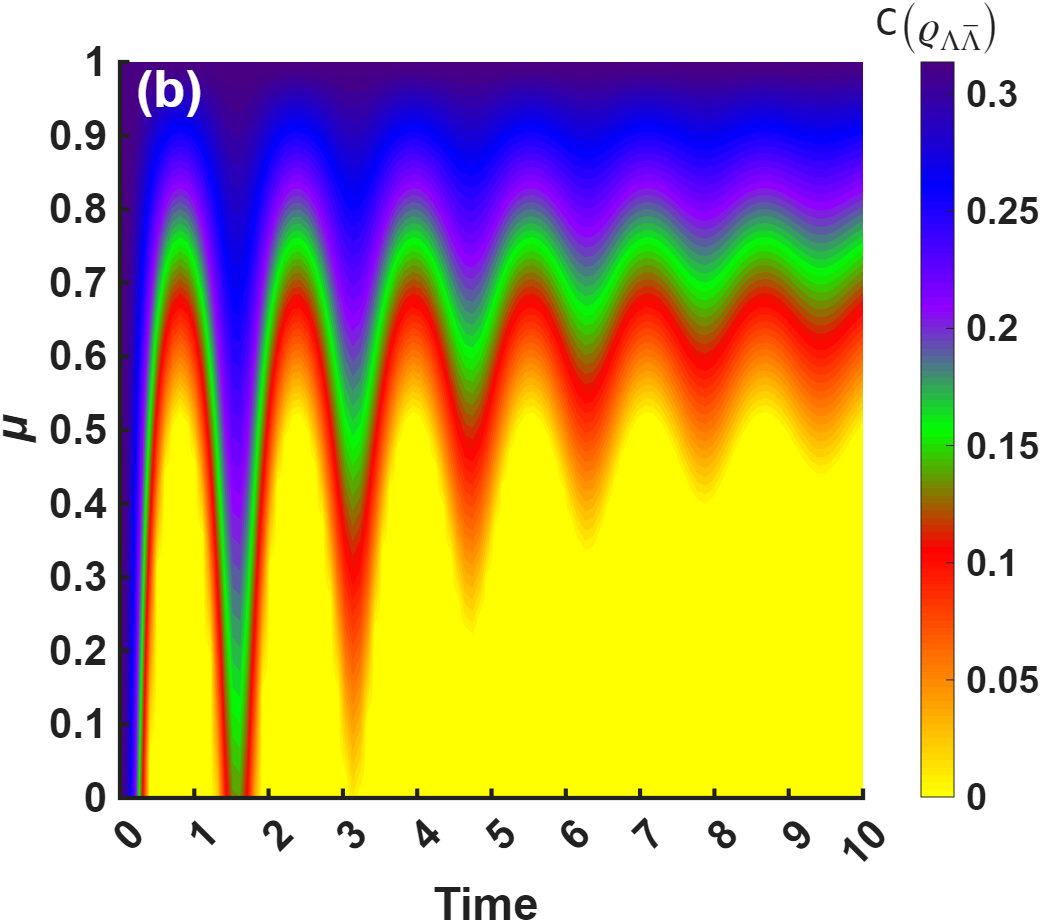}
\caption{Evolution of $\mathtt{C}\big(\varrho_{\Lambda\bar{\Lambda}}\big)$ under different environmental memory effects: (a) the Markovian case ($\tau = 0.2$) and (b) the non-Markovian case ($\tau = 20$), with mass corrections omitted and $\varphi = \pi/2$.}
\label{fig:Mac1}
\end{figure}
\begin{figure}[!h]
\includegraphics[scale=0.4]{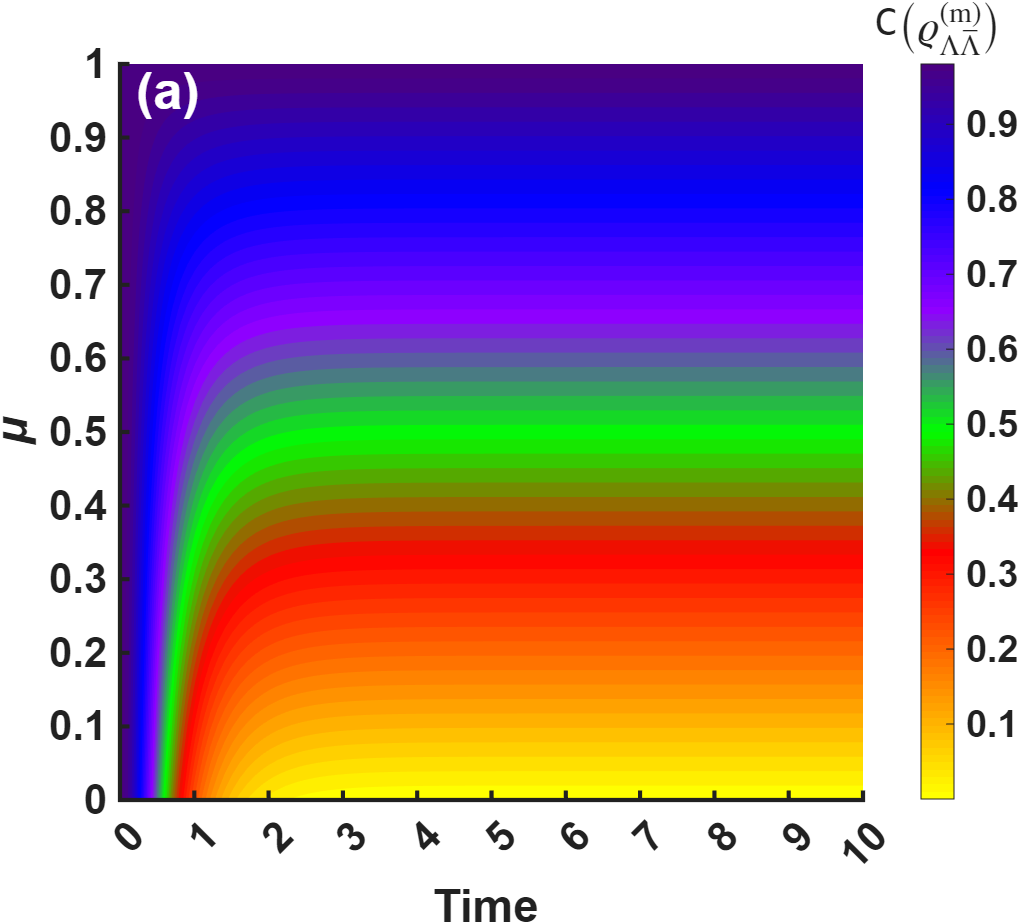}
\includegraphics[scale=0.4]{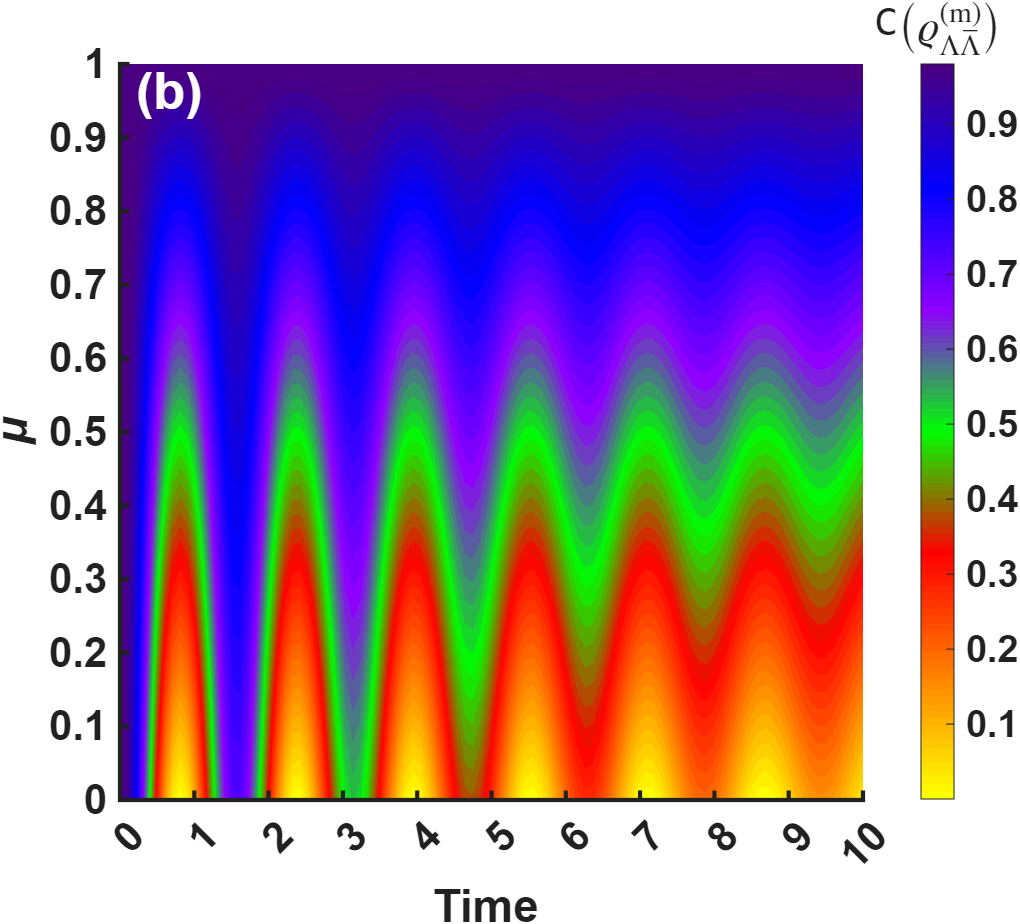}
\caption{The effect of mass corrections on the time evolution of the concurrence $\mathtt{C}\big(\varrho^{\rm (m)}_{\Lambda\bar{\Lambda}}\big)$ is illustrated for $\varphi=\pi/2$ in (a) the Markovian regime ($\tau = 0.2$) and (b) the non-Markovian regime ($\tau = 5$).}
\label{fig:Mac2}
\end{figure}

Figures~\ref{fig:Mac1}(a-b) present the evolution of the concurrence $\mathtt{C}\big(\varrho_{\Lambda\bar{\Lambda}}\big)$ as a function of time and the classical correlation parameter $\mu$ in the Markovian regime. The results clearly indicate that increasing $\mu$ slows down the degradation of entanglement, revealing that stronger classical correlations in the environment partially protect the quantum correlations of the system against decoherence. After a certain evolution time, the entanglement approaches a stationary regime, as shown in Fig.~\ref{fig:Mac1}(a). Furthermore, the steady-state value of concurrence increases with $\mu$, highlighting the constructive role played by correlated environmental noise in preserving entanglement.

The influence of classical correlations on the entanglement dynamics is further illustrated in Fig.~\ref{fig:Mac2}(a-b). As the parameter $\mu$ increases, the oscillatory behavior of $\mathtt{C}\big(\varrho_{\Lambda\bar{\Lambda}}\big)$ becomes progressively damped, while the overall decay rate of concurrence is significantly reduced. This behavior reflects the interplay between classical and quantum correlations: stronger classical correlations in the noise process effectively suppress the loss of quantum coherence and favor the persistence of entanglement in the $\Lambda\bar{\Lambda}$ system.

\begin{figure}[!h]
\includegraphics[scale=0.4]{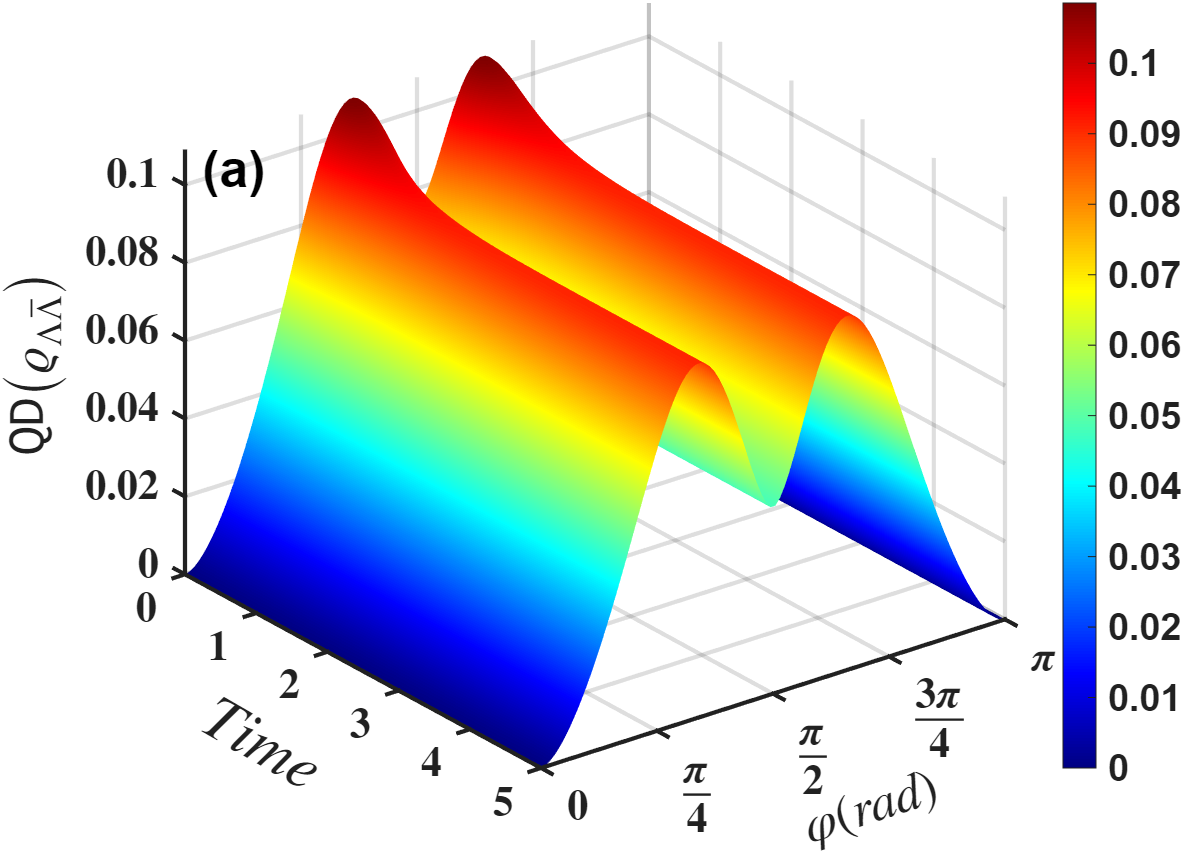}
\includegraphics[scale=0.4]{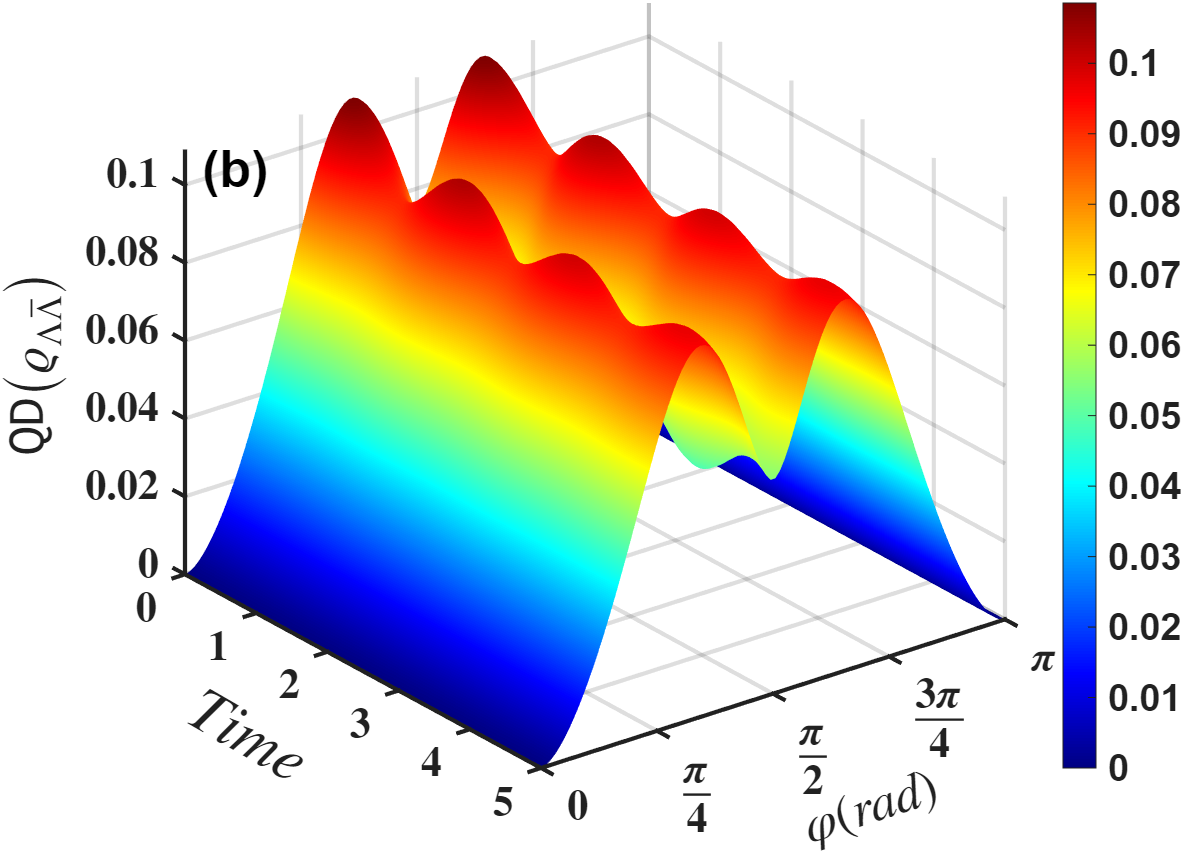}
\caption{Evolution of $\mathtt{QD}\big(\varrho_{\Lambda\bar{\Lambda}}\big)$ under different environmental memory effects: (a) the Markovian case ($\tau = 0.2$) and (b) the non-Markovian case ($\tau = 5$), with mass corrections omitted and $\mu = 0.1$.}
\label{fig:d1}
\end{figure}

\begin{figure}[!h]
\includegraphics[scale=0.4]{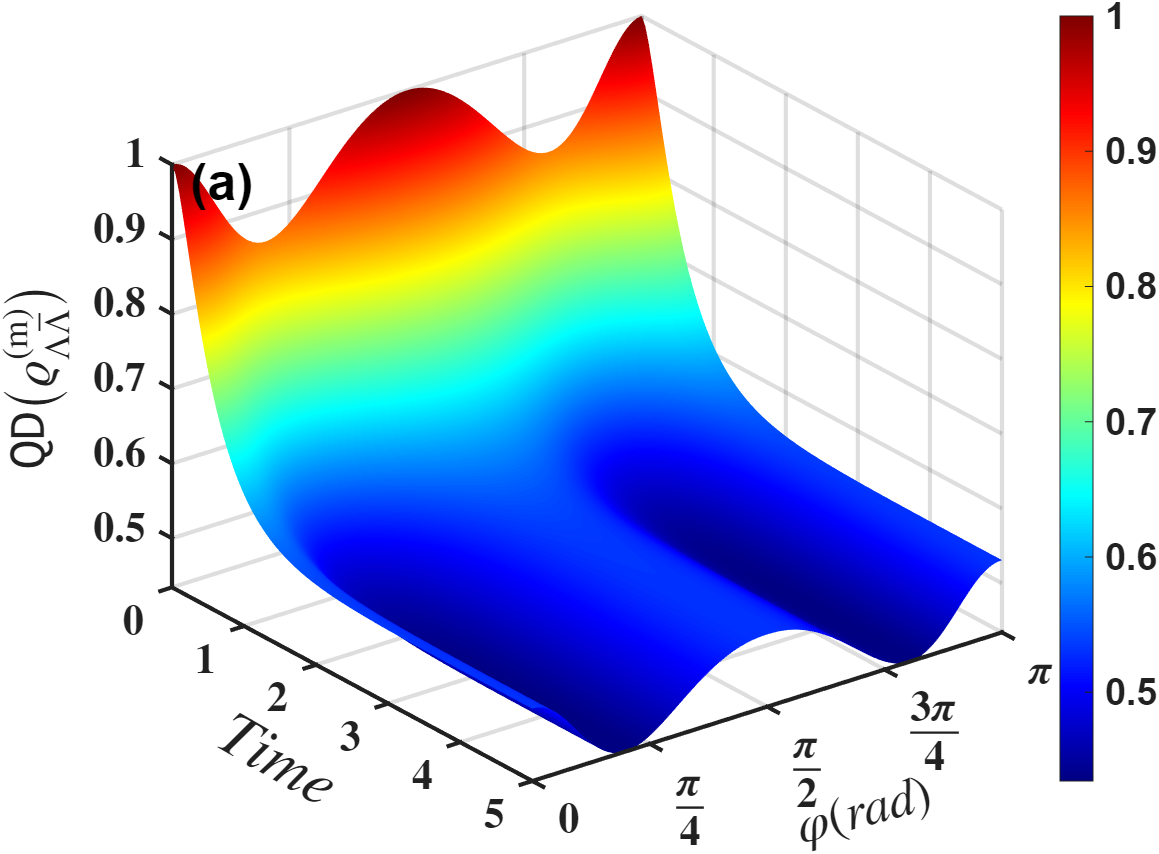}
\includegraphics[scale=0.4]{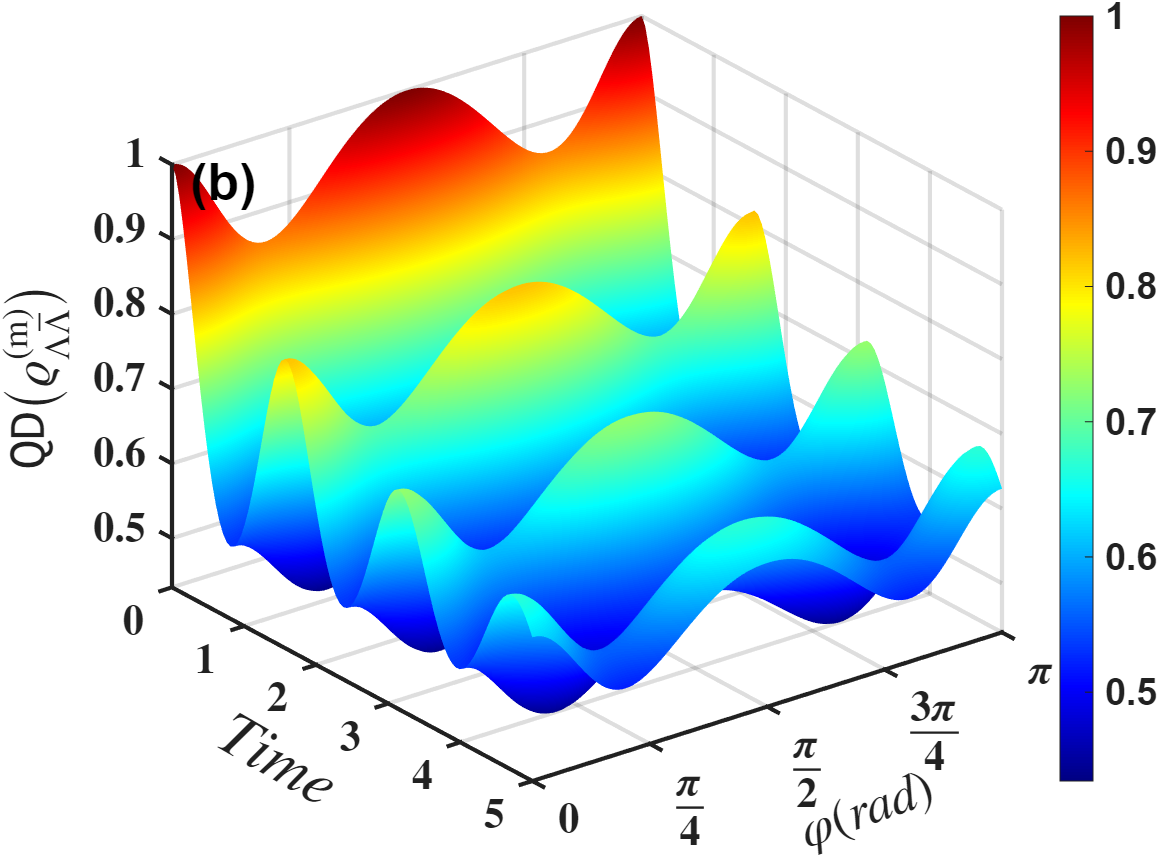}
\caption{The effect of mass corrections on the time evolution of the quantum discord $\mathtt{QD}\big(\varrho^{\rm (m)}_{\Lambda\bar{\Lambda}}\big)$ is illustrated for $\mu = 0.8$ in (a) the Markovian regime ($\tau = 0.2$) and (b) the non-Markovian regime ($\tau = 5$).}
\label{fig:d2}
\end{figure}
The dynamics of the QD, denoted as $\mathtt{QD}\big(\varrho_{\Lambda\bar{\Lambda}}\big)$, are illustrated in Fig. \ref{fig:d1}(a) and Fig. \ref{fig:d1}(b) for the Markovian and non-Markovian regimes, respectively. To ensure a consistent comparison across different quantum correlation measures, the QD calculations utilize the same parameter space previously employed for the analysis of Bell non-locality, quantum steering and concurrence. In the Markovian limit [Fig. \ref{fig:d1}(a)], the mass-independent the QD of the baryon-antibaryon system is symmetric with respect to $\varphi=\pi/2$. In the whole range $\varphi\in[0,\pi]$, the QD of spin states of $\Lambda\bar{\Lambda}$ is non-zero. Additionally,
unlike the Bell non-locality, quantum steering and quantum concurrence, the maximum quantum
discord is not necessarily located at $\varphi=\pi/2$, the discord has two peaks away from $\varphi=\pi/2$. As time evolves, a monotonic decay in $\mathtt{QD}\big(\varrho_{\Lambda\bar{\Lambda}}\big)$ is observed, signifying the gradual erosion of quantum correlations driven by environmental decoherence. 

In the Markovian limit (Fig.~\ref{fig:d1}(a)), QD gradually decreases over time, reflecting the irreversible loss of quantum correlations due to environmental dissipation. In the non-Markovian regime (Fig.~\ref{fig:d1}(b)), the dynamics of QD deviate significantly from a monotonic decay. Instead, QD undergoes oscillatory behavior with partial revivals over time, characteristic of memory effects in the environment. These oscillations are a manifestation of information backflow, whereby the environment temporarily restores quantum correlations to the $\Lambda\bar{\Lambda}$ pair.

In the Markovian limit [Fig. \ref{fig:d2}(a)], the mass-dependent QD exhibits a pronounced angular symmetry, with global maxima occurring at $\varphi = {0, \pi/2, \pi}$. As time progresses, a monotonic decay of $\mathtt{QD}\big(\varrho_{\Lambda\bar{\Lambda}}\big)$ is observed, reflecting the gradual erosion of quantum correlations driven by environmental decoherence. Conversely, Fig. \ref{fig:d2}(b) illustrates the QD evolution within the non-Markovian regime as a function of both time and scattering angle $\varphi$. In this case, the temporal profile of $\mathtt{QD}\big(\varrho_{\Lambda\bar{\Lambda}}\big)$ deviates from monotonic decay, displaying instead periodic oscillations that constitute a distinct signature of non-Markovianity. The underlying memory effects induce a non-monotonic temporal behavior, signifying a backflow of information from the environment to the system.
\begin{figure}[!h]
\includegraphics[scale=0.4]{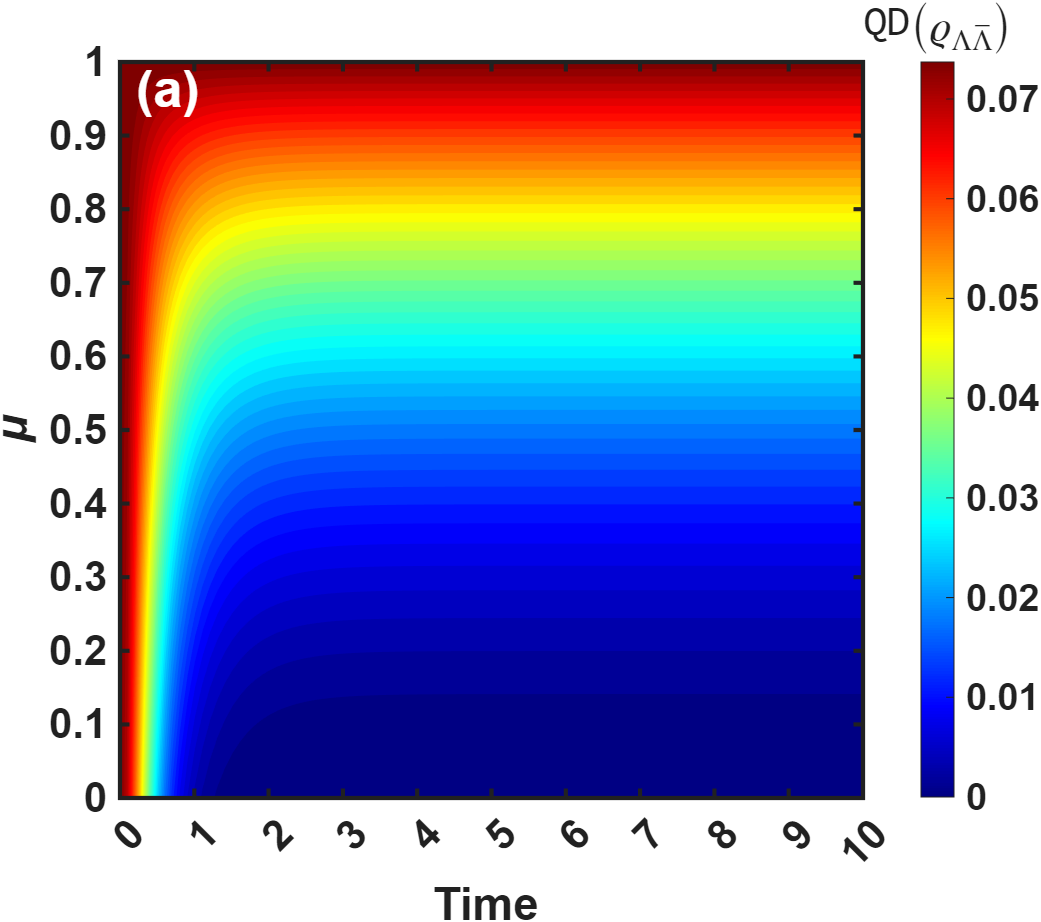}
\includegraphics[scale=0.4]{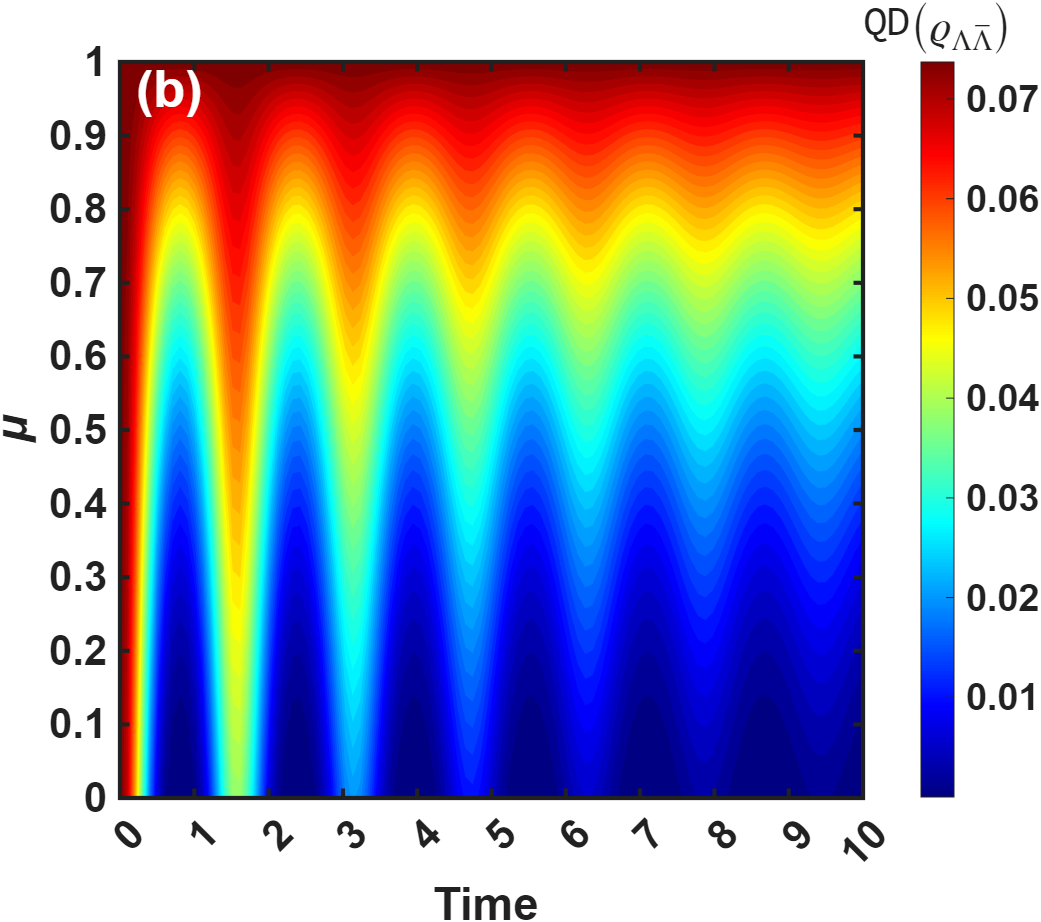}
\caption{Evolution of $\mathtt{QD}\big(\varrho_{\Lambda\bar{\Lambda}}\big)$ under different environmental memory effects: (a) the Markovian case ($\tau = 0.2$) and (b) the non-Markovian case ($\tau = 5$), with mass corrections omitted and $\varphi = \pi/2$.}
\label{fig:d3}
\end{figure}

\begin{figure}[!h]
\includegraphics[scale=0.4]{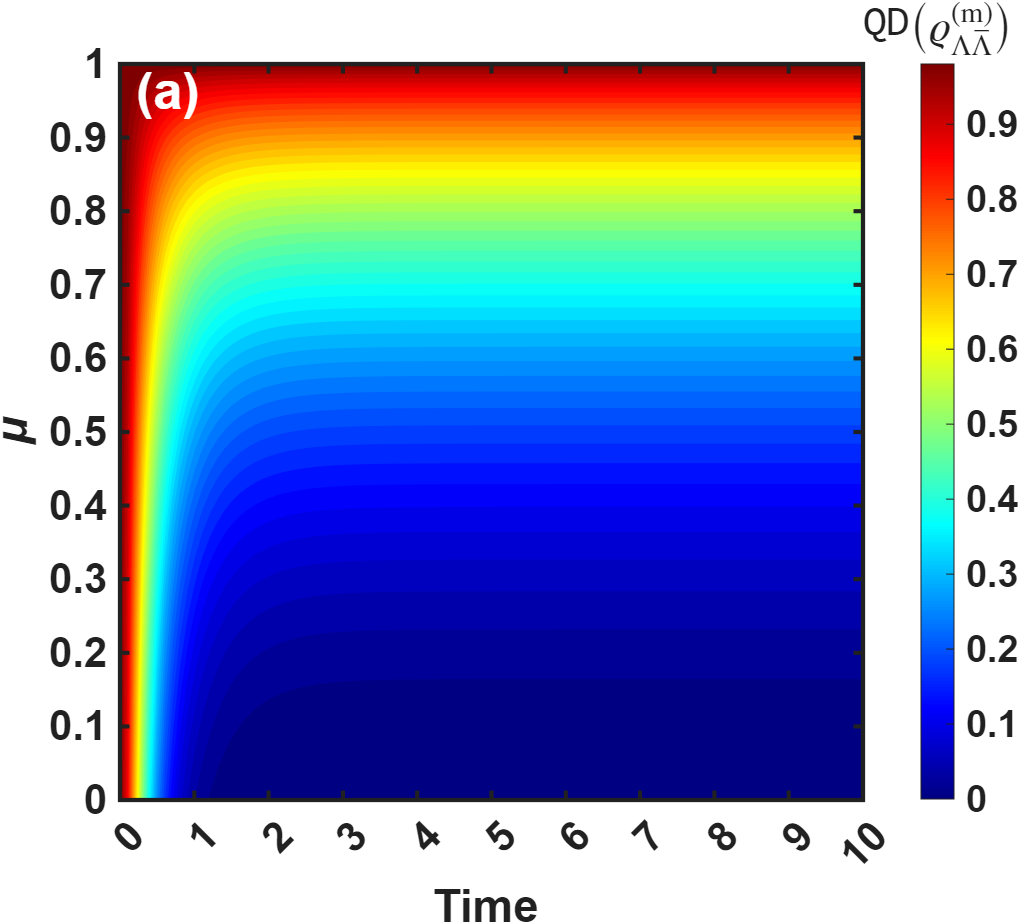}
\includegraphics[scale=0.4]{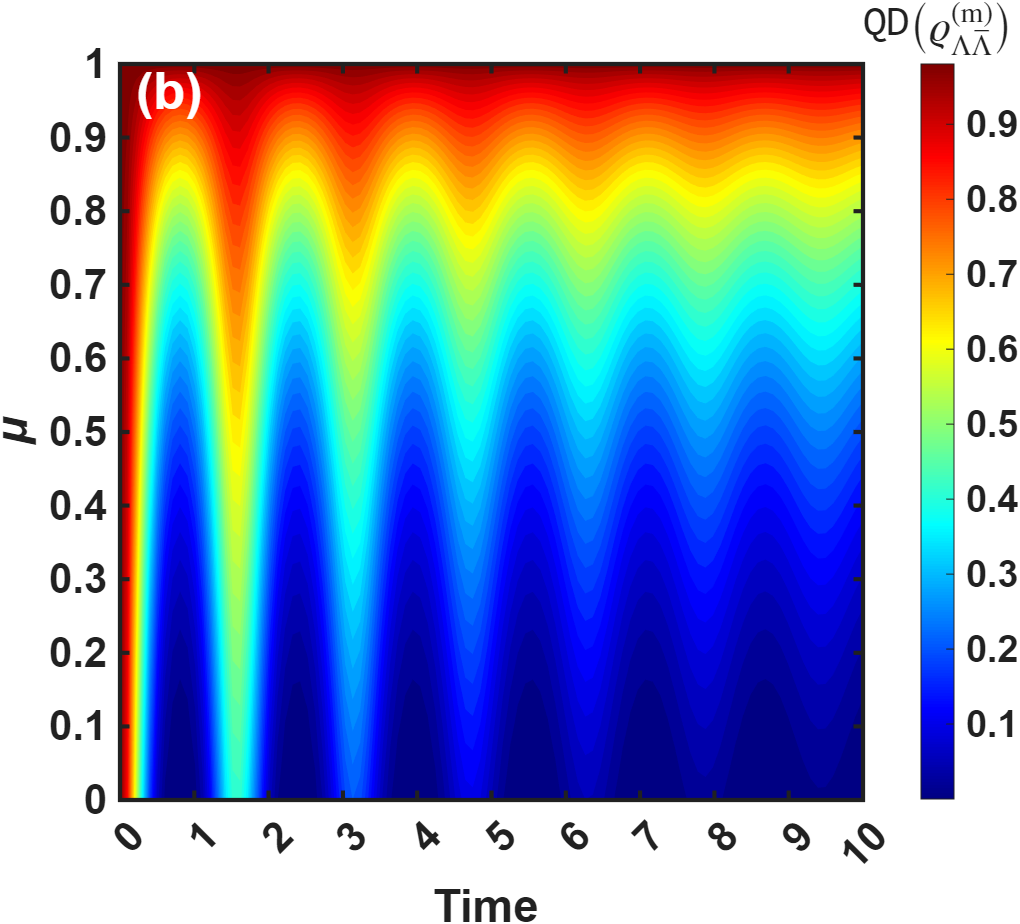}
\caption{The effect of mass corrections on the time evolution of the QD $\mathtt{QD}\big(\varrho^{\rm (m)}_{\Lambda\bar{\Lambda}}\big)$ is illustrated for $\varphi=\pi/2$ in (a) the Markovian regime ($\tau = 0.2$) and (b) the non-Markovian regime ($\tau = 5$).}
\label{fig:d4}
\end{figure}

Figures~\ref{fig:d3} and \ref{fig:d4} illustrate the impact of classical correlations in the dephasing channel on the evolution of QD. As the classical correlation parameter $\mu$ increases, the decay of QD becomes progressively slower, demonstrating that stronger classical correlations enhance the preservation of quantum correlations. In the non-Markovian regime, memory effects induce non-exponential decay and richer dynamical features, including oscillations and partial revivals of discord. For $\mu$ approaching unity, QD remains remarkably robust against decoherence in both Markovian and non-Markovian regimes, highlighting the protective role of classical correlations combined with environmental memory effects.
\section{Coclusion}\label{sec:6}

In conclusion, we have derived the bipartite density matrix for the $e^{+}e^{-} \rightarrow J/\psi \rightarrow \Lambda \bar{\Lambda}$ process at BESIII. We have investigated the impact of mass corrections and memory effects (within Markovian and non-Markovian regimes) on quantum correlations even beyond entanglement. By employing a comprehensive set of quantifiers-steering inequalities, concurrence, quantum discord, and Bell-type inequalities-we have systematically characterized the quantum resources of the $\Lambda\bar{\Lambda}$ system. Our analysis has revealed a pronounced angular dependence, with maximal correlations at $\varphi=\pi/2$, and has highlighted the significant impact of baryon mass corrections, which have enhanced Bell violations and modified angular distributions. We have shown that classical correlations and non-Markovian memory effects have slowed decoherence, stabilized quantum resources, and induced partial revivals. Overall, our results have confirmed the theoretical hierarchy of quantum correlations-$\text{Bell non-locality}\subset\text{Steering}\subset\text{Entanglement}\subset\text{Discord}$-and have provided new insights for the control and engineering of quantum correlations in high-energy particle processes, bridging the interface between quantum information theory and collider physics.

\bibliography{sample}
\end{document}